\documentclass[10pt]{article}

\usepackage{graphicx}
\usepackage[english]{babel}
\usepackage{amsmath}
\usepackage{epstopdf}
\usepackage{authblk}

\newcommand{\bsm}{\boldsymbol}

\newcommand{\bdot}{\boldsymbol{\cdot}}

\begin{document}

%%%%%%%%%%%%%%%%%%%%%%%%%%%%%%%%%%%%%%%%%%%%%%%%%%%%%%%%%%%%%%%%%%%%%%%%%%%%%%%%
% Title
%%%%%%%%%%%%%%%%%%%%%%%%%%%%%%%%%%%%%%%%%%%%%%%%%%%%%%%%%%%%%%%%%%%%%%%%%%%%%%%%

\title{Quantum modeling of ultrafast photoinduced charge separation}

\author[1]{Carlo Andrea Rozzi\thanks{carloandrea.rozzi@nano.cnr.it}}
\author[1]{Filippo Troiani}
\author[2]{Ivano Tavernelli}
\affil[1]{Istituto Nanoscienze -- Consiglio Nazionale delle Ricerche, Modena, Italy}
\affil[2]{IBM Research, Zurich Research Laboratory, Zurich, Switzerland}

\maketitle

%%%%%%%%%%%%%%%%%%%%%%%%%%%%%%%%%%%%%%%%%%%%%%%%%%%%%%%%%%%%%%%%%%%%%%%%%%%%%%%%
% Abstract
%%%%%%%%%%%%%%%%%%%%%%%%%%%%%%%%%%%%%%%%%%%%%%%%%%%%%%%%%%%%%%%%%%%%%%%%%%%%%%%%

\begin{abstract}
Phenomena involving electron transfer are ubiquitous in nature, photosynthesis and enzymes or protein activity being prominent examples. Their deep understanding thus represents a mandatory scientific goal. Moreover, controlling the separation of photogenerated charges is a crucial prerequisite in many applicative contexts, including quantum electronics, photo-electrochemical water splitting, photocatalytic dye degradation, and energy conversion. In particular, photoinduced charge separation is the pivotal step driving the storage of sun light into electrical or chemical energy. If properly mastered, these processes may also allow us to achieve a better command of information storage at the nanoscale, as required for the development of molecular electronics, optical switching, or quantum technologies, amongst others.

In this Topical review we survey recent progress in the understanding of ultrafast charge separation from photoexcited states. We report the state-of-the-art of the observation and theoretical description of charge separation phenomena in the ultrafast regime mainly focusing on molecular- and nano-sized solar energy conversion systems. In particular, we examine different proposed mechanisms driving ultrafast charge dynamics, with particular regard to the role of quantum coherence and electron-nuclear coupling, and link experimental observations to theoretical approaches based either on model Hamiltonians or on first principles simulations.
\end{abstract}

%%%%%%%%%%%%%%%%%%%%%%%%%%%%%%%%%%%%%%%%%%%%%%%%%%%%%%%%%%%%%%%%%%%%%%%%%%%%%%%%
\section{Introduction}\label{sec:outline}
%%%%%%%%%%%%%%%%%%%%%%%%%%%%%%%%%%%%%%%%%%%%%%%%%%%%%%%%%%%%%%%%%%%%%%%%%%%%%%%%

% General and applications
The displacement of charge due to light absorption occurs in a variety of both natural and artificial environments, ranging from complex macromolecular biological systems to solid state inorganic junctions and hybrid inorganic nanoparticle -- organic molecule -- liquid solvent interfaces. 

Photoinduced electron transfer phenomena make up the foundations of vital natural functions such as photosynthesis, vision, and DNA damage repair \cite{Blankenship_2002,Gray_1996a,Weber_2005a}. The understanding of these reactions is {\em per se} a paramount scientific goal. The ability of optimizing and controlling them offers the chance to solve high impact technological and social challenges in applications such as solar energy conversion, artificial photosynthesis, photocatalysis, molecular electronics, photodynamic therapy, etc. \cite{Balzani_2001a,Petty_2007}. A description of these phenomena at a fully microscopic quantum level has been only partially reached, and its achievement remains a major scientific challenge.

% ET time scales
Conceptually, photoinduced charge separation shares a similar scheme across different systems. Electron dynamics in molecules and solids proceeds on an attosecond time scale, while nuclear dynamics can involve low-frequency vibrations and slow displacements that may take hundreds of picoseconds to complete. However, this simple picture, which founds the Born-Oppenheimer approximation, can be very deceiving. Part of the complexity of the problem consists in the fact that electron transfers can in fact occur across a wide range of time scales. On the one hand, they can be made slower by decreasing the coupling between donors and acceptors (e.g. in the mitochondrial inner membrane they can take up to tens of ms \cite{Trouillard_2011}); on the other hand, non-Born-Oppenheimer dynamics can dramatically influence the charge separation in times as short as few fs (see section \ref{sec:proto} for a detailed discussion).

Considering a statistical ensemble of mutually independent transfer reactions between identical replicas of donors D and acceptors A, the overall energetics can be understood by looking at panel a) of Figure \ref{fig:marcus}.

The absorption of a photon induces a sudden modification of the electron distribution, which is energetically unstable at the reactants nuclear arrangement D$^*$A. To accommodate this change, a reconfiguration occurs along suitable reaction coordinates (only one for simplicity in the figure). Eventually, the product state --i.e. a charge-separated state-- is reached at the reaction coordinate D$^+$A$^-$. The reaction proceeds bypassing an energy barrier, and the reaction rate depends on the electronic coupling and energy gaps of the DA system. It is also influenced by the electronic relaxation due to the surrounding medium (e.g. the solvent in the case of a molecule, or whatever is at the boundaries in the case of more complex systems).

Even considering the initial photon absorption as instantaneous, we cannot neglect the fact that the nuclear rearrangement may involve both high- and low-frequency vibrational modes, and that the environmental contribution can have both a fast inertial and a slow diffusive-like response. In a solid-state embedding the role of vibrational modes is taken by phonons, and other quasiparticles, such as polarons, and can arise on the same energy scale.

To further complicate the picture, long-range charge movements between D and A sites may proceed in a direct (also called coherent, or superexchange) process or via a sequence of incoherent hops, depending on the details of the D-A electronic and of  electron-nuclei couplings.

This is not the end of the story yet. Panel b) of figure 1 shows that a number of energy loss and electron-hole recombination channels impair charge separation. Among them, internal conversion, inter-system crossing, and fluorescence. In order to be efficient, charge separation must dynamically compete with these processes. This is why the understanding of charge separation dynamics is bound to the broader problem of efficiency. The way this is obtained in nature is by a suitable organization of the environment.

Recent advances in experimental techniques are achieving an unprecedented spatial and temporal resolution, and are moving towards the microscopic control of charge, energy and information flows. In particular, the development of time-resolved spectroscopic methods has allowed researchers to shed new light on the ultrafast dynamics governing charge separation processes. These techniques have shifted the core of experimental methods from classical electrochemical measurement, addressing energy differences between the initial and final states, to time-resolved methods, also capable of monitoring the evolution of transient states. Under this respect, many observables may provide information, such as time-resolved photocurrent and transient optical absorption. Moreover, more sophisticated settings employing a sequence of light pulses may be employed to map the correlations between excitations at different energies (2D spectroscopy). These methods will be briefly reviewed in section \ref{sec:exp}.

These advances demand the basic understanding of the steps driving the different processes described in figure \ref{fig:marcus}. In order to possibly obtain a full control of them, we can no longer rest on the ensemble averaged equilibrium description of charge transfer reactions. The development of fully quantum mechanical schemes is needed, both to rationalize the precise space and time scales of the phenomena and to possibly steer the synthesis and engineering of new efficient materials supporting them.

The achievement of such a comprehensive vision is particularly challenging theoreticians for several reasons. First, photoinduced charge separation always proceeds through excited states. Therefore simple views based on adiabatic connections between equilibrium states cannot disclose the entire dynamics of charge or energy flow. Second, charge separation can proceed through a variety of qualitatively diverse channels, having entirely different length and time scales, sometimes deeply interleaved. In particular, the coupling between electronic and vibrational degrees of freedom requires that a proper quantum-mechanical, or at least a semi-classical treatment of the problem is formulated. Last, due to the variety of environments in which charge separation occurs, efficient computational schemes must be designed to tackle the problem in its generality, and in a way suitable to be applied at sparse scales and for diverse systems.

\section{Outlook}

% Scope of the review
Given the vast diversity of charge transfer phenomena, the scope of this review is limited to ultrafast photoinduced charge separation in molecular and supramolecular systems (especially as components of natural and artificial light-harvesting complexes) and to nanostructured materials for solar energy conversion.

% Summary of the paper
This work is structured as follows: sections \ref{sec:intro}--\ref{sec:cur} are dedicated to sketch the basic features of photoinduced ultrafast charge separation and outline the experimental observables describing it. This is not meant to be a review of experimental techniques, but aims to introduce the variety of problems posed in several classes of materials,  and to provide information about the physical observables to be directly compared to theoretical models and simulations. In sections \ref{sec:quantcoh}--\ref{sec:thspect} we examine in depth the concepts of coherence and entanglement, and review theoretical methods based on the identification of few, relevant degrees of freedom, where charge separation is described in the framework of open quantum systems. Sections \ref{sec:QD}--\ref{sec:wfmeth} are dedicated to review first principles and semi-empirical time-dependent atomistic methods adopted to solve the coupled electronic and molecular dynamics.

%%%%%%%%%%%%%%%%%%%%%%%%%%%%%%%%%%%%%%%%%%%%%%%%%%%%%%%%%%%%%%%%%%%%%%%%%%%%%%%%
\section{General features of photoinduced charge separation}\label{sec:intro}
%%%%%%%%%%%%%%%%%%%%%%%%%%%%%%%%%%%%%%%%%%%%%%%%%%%%%%%%%%%%%%%%%%%%%%%%%%%%%%%%

% General description
Photoinduced charge separation is a process by which, upon absorption of photons, a redox reaction occurs in the excited state of a system. The final products of the excited state reaction are spatially separated and unbound charges in molecular species, or free charge carriers in bulk solids.

When charge separation occurs between two molecular species ---a donor D and an acceptor A--- it is often described in a multi-step framework, proceeding through a number of intermediate states, as summarized in table \ref{tab:steps}.
\begin{table}[h]
  \begin{center}
  \begin{tabular}{|p{0.45\textwidth}|}
    \hline
    {\bf 1)} the donor excitation $D + A \to D^* + A$ \\
    {\bf 2)} the excitation induced delocalization $D^* + A \to (DA)^*$ \\
    {\bf 3)} the charge transfer (CT) leading to the formation of the radical pairs $(DA)^* \to (D^{+\bdot}A^{-\bdot})$ \\
    {\bf 4)} the charge separation $(D^{+\bdot}A^{-\bdot}) \to D^+ + A^-$. \\
    \hline
  \end{tabular}
  \caption{Common multi-step description of charge separation.}
  \label{tab:steps}
  \end{center}
\end{table}

More generally speaking, however, in step 1) the reaction could be initiated by acceptor excitation, instead of donor excitation; the product of step 3), the CT state, may also be reached without the need of step 1) if indirect excitons can be generated at the interface, i.e. if the overlap of donor and acceptor orbitals allows for it. 

Speaking about terminology, note that in the literature, depending on the context, the product of step 3) may also be referred to as geminate pairs, bound polarons or exciplexes. Also note that in some cases in the literature ``charge transfer" and ``charge separation" are considered synonymous, which is not the case in the present review.

% Easy energetics
From a mere energetical point of view, an electron transfer reaction can proceed if the ionization energy $I_{D^*}$ {\em of the excited state} of the donor is smaller than the sum of the acceptor electronic affinity $A_A$, and the total electrostatic energy $U_{DA}$ between the donor and the acceptor: $I_{D^*} < A_A +U_{DA}$. $U_{DA}$ must also include polarization effects induced by the modification in the nuclear configuration associated with the reaction. However, the simplicity in the energetics hides the details of the process, and a description exclusively based on energetics of initial and final states fails to capture the richness of the phenomenon.

% Difficulty in understanding organic solar cells and how ultrafast can help
Even restricting the field, for instance, to the case of organic solar cells, and assuming that the energetics of reactants and products is ultimately known, challenges still exist in understanding the basic charge separation mechanism: the low dielectric constant of organic materials implies that the screening of coulombic attraction is weak, and therefore, in the excited state, photoexcited electron and hole pairs are generated with large binding energy (typically in the range of 0.1-0.5 eV \cite{Alvarado_1998a}, one order of magnitude larger than the thermal energy at room temperature). The driving force necessary to overcome this energy may come from several independent isoenergetic channels. It is clear that much richer information can be gained by examining the kinetics and the dynamics of the reactions.

% ET kinetics
For what concerns the kinetics, the traditional reference point is Marcus' theory of electron transfer (for a review see \cite{Marcus_1993a}). According to such theory, the rate constant for an electron transfer occurring at temperature $T$ is linked to the electronic coupling $V$ between the donor and the acceptor and to the activation energy of the reaction $\Delta G^*$ by the relation 
\begin{equation}
  k_{CT} = |V|^2 \sqrt{\frac{4\pi}{h^2k_BT\lambda}} \exp\left(-\frac{\Delta G^*}{k_BT}\right).
\end{equation}
$\Delta G^*$ depends both on the Gibbs free energy change of the reaction $\Delta G^0_{CT}$, and on the reorganization energy $\lambda$, which is the energy change associated with a change in the nuclear configuration of the system due to the combined effects of atomic motion and solvation.

Assuming parabolic shapes for the diabatic potential energy surfaces as functions of a (one-dimensional) reaction coordinate, we get
\begin{equation}
  \Delta G^* = \frac{(\Delta G^0+\lambda)^2}{4\lambda}.
\end{equation}
Within this framework, it is easy to see that, as the free energy difference $\Delta G^0_{CT}$ becomes more negative, the transfer rate increases (normal region). When $-\Delta G^0_{CT} = \lambda$, $k_{CT}$ reaches a maximum, and any further decrease of $\Delta G^0_{CT}$ causes a decrease of the transfer rate (inverted region).

% Predictions of Marcus' model
The very existence of this path (Marcus' parabola) and the central role of the reorganization energy in the process suggest that, by carefully optimizing the value of $\lambda$ in a given environment, it is possible to tune the kinetics in such a way that the electron transfer is much faster than the inverse back transfer. This kind of optimization is believed to happen in natural phenomena such as photosynthesis, in order to produce long-lived charge separated states.

% Limits of Marcus' model
Although extremely successful and predictive, this model rests on several crucial simplifications: the potential energy surfaces are parabola; all the nuclear modes are classical; Fermi's golden rule holds for electronic vertical transitions; redistribution of vibrational energy in the excited state occurs faster than the charge transfer, i.e. charge transfer is initiated from the equilibrium population of vibrational modes. This means that classical Marcus' theory is not generally suited to address the coherent electron transfer regime that can arise after the interaction with femtosecond or attosecond laser pulses has triggered the vibronic wavepacket motion, due to non-adiabatic electron-nuclei coupling. In this regime, the CT reactions are quite sensitive to non-equilibrium dynamical effects involving nuclear motions of the reactants and of the environment \cite{Barbara_1996a}. More suitable schemes to address this regime will be reviewed in the next Sections.

% Meaning of ultrafast
Therefore, studying the dynamics of charge separation proves to be even more difficult that studying its kinetics, because of the interplay between internal (electronic and nuclear) degrees of freedom and the external environment. Fortunately, ultrafast optical spectroscopy can substantially contribute to the identification of the different excitations that appear in the same energy range by differentiating them according to the respective decay times, becoming therefore a tool of choice in the investigation of photoinduced charge separation.

One final word is about the meaning of the adjective {\em ultrafast}, which actually has followed the progress of time-resolved experimental techniques. In the literature it is found to refer to a wide variety of time scales. In the present review we will mostly focus on the sub-ps regime, with a particular eye for sub-100 fs phenomena.

%%%%%%%%%%%%%%%%%%%%%%%%%%%%%%%%%%%%%%%%%%%%%%%%%%%%%%%%%%%%%%%%%%%%%%%%%%%%%%%%
\section{Experimental techniques for the study of photoinduced charge separation}\label{sec:exp}
%%%%%%%%%%%%%%%%%%%%%%%%%%%%%%%%%%%%%%%%%%%%%%%%%%%%%%%%%%%%%%%%%%%%%%%%%%%%%%%%

From an experimental point of view, many techniques can be employed to provide evidence of charge separation and monitor its time evolution. This section is not meant as a technical experimental review, but rather as a way to introduce the observed phenomenology of ultrafast charge separation. In later sections, we will discuss to what extent these observations can be theoretically reproduced or predicted.

Early methods based on photoelectrochemical techniques consisted in measuring steady state voltage-current curves under illumination \cite{Becquerel_1839}. These approaches are now empowered by the possibility of using sequences of ultrafast laser pulses to create excited states in a controlled way and probe their evolution in time. This technique is particularly valuable when assessing the performance of photovoltaic devices, since photocurrent is the primary factor of merit in this case.

Looking at figure \ref{fig:marcus}, we realize that charge transfer reactions cause, as they proceed, a set of changes in the excited states (both electronic and nuclear) of the material. Since the spacing and nature of the energy levels is modified as long as the charge flows, it is possible to extract information about the reaction using spectroscopical techniques, i.e. by observing energy differences between one or more excited states and the ground states of reactants and products. The time-dependent differences between excitation spectra in dark and under illumination can reveal how the reaction is proceeding. These spectra are obtained by using a pump pulse to populate excited states of the system, optionally followed by one or more control pulses, and terminated by a weak probe pulse to observe the state at given time delays.

A simple classification of spectroscopic techniques can be done according to the observed quantities: optical spectroscopies measure the amount of light absorbed and/or emitted by the sample upon photoexcitation, and can monitor both electronic excitations, in the visible or ultraviolet bands, and nuclear vibrations, via infrared and Raman spectroscopy. They are very well suited to study thin films and molecular samples in solution.

Besides the obvious advantage of being a time-resolved technique, the differential absorption spectrum contains both negative and positive contributions from various processes: ground-state bleaching, stimulated emission (both negative), and product absorption (positive). This allows one to investigate the evolution of non-emissive and dark states that cannot be seen in time-resolved fluorescence or steady-state absorption.

A different approach is required when one needs information not only on the excitation energies, but also on the correlations between the different excitations. This is pretty much desired when investigating the nature of coherences. In this case, one needs bidimensional (2D) spectra, i.e. maps having two independent excitation and probe frequencies on the two axes. A  sequence of three noncollinear pulses is used to record 2D spectra, where disorder and different molecular orientations give rise to inhomogeneous broadening (see also the discussion in section \ref{sec:nonlinspec}).

Electronic spectroscopies observe outgoing electrons, i.e. either the generated photocurrent flowing through electrodes or the photoemitted electrons. In particular, photoemission is a surface-sensitive technique, particularly useful when studying molecular absorbates on crystal surfaces. Furthermore, it can be angle-resolved, providing detailed maps of the distribution of electrons in momentum space, and it can also extract electrons from core levels.

% Optical absorption
{\bf Watching absorbed photons.} Absorption and emission spectra can reveal charge-transfer states as below band features, because CT states lie below the band gap of the donor and acceptor components. However, if the donor and acceptor frontier wavefunctions do not overlap significantly, as is often the case, CT excitations have a very small cross section, and can be only revealed by enhancement techniques such as photothermal deflection spectroscopy \cite{Clarke_2010a}.

% Transient absorption
The time-dependent version of steady state optical absorption, transient absorption, is able to probe excited states and the coupling between them with unprecedented (10 fs) resolution \cite{Cerullo_2002a}. Transient absorption allows one to monitor the formation of ion radicals in polymers as the stimulated emission band of the photoexcited singlet exciton evolves into the characteristic photo-induced absorption band of the charged species \cite{Brabec_2001a}. Transition absorption also allows us to track the evolution of non-emissive species such as polarons and triplet excitons. This powerful technique is most profitably complemented by theoretical modeling and simulations, because different excited states usually contribute to photoinduced absorption and stimulated emission in the same spectral band, and their identification may be impossible by just looking at the transient spectra. However transient absorption has shown that CT can occur on an ultrafast time scale (about 50 fs in polymer-fullerene blends) \cite{Brabec_2001a,Grancini_2012a} and this is the most relevant fact motivating this review.

% mid-IR and near-IR spectroscopy
In some cases, formation and recombination of CT excitons can also be observed by looking at solvatochromic and Stark shifts in vibrational frequencies. For example, in polymer-PCBM aggregates the displacement of electronic density of the charge-separated state with respect to the excitonic and ground state determines a polarization of the environment which in turn shifts the vibrational frequencies of the acceptor \cite{Barbour_2007a}.

When electrons are transferred from molecular adsorbates to nanostructured or bulk semiconductor acceptors, as is the case in dye-sensitized solar cells, the arrival of electrons inside the semiconductor can be probed by using infrared femtosecond spectroscopy, since free carrier absorption, intraband transitions, and trap state absorption in semiconductors all occur in the mid-infrared region of the spectrum \cite{Asbury_2001a}.

% 2D spectroscopy
Time-resolved optical spectroscopy is capable of providing a huge deal of information regarding the energetics and the dynamics of optical transitions, and oscillatory transient signals may be taken as indicators that underlying coherences (either electronic or vibrational) are present. However, as mentioned above, identifying the nature of excitations and coherences often requires assistance from theory and modeling. Two-dimensional spectroscopy, in its several flavors \cite{Jonas_2003a}, allows us to correlate excitation frequencies in different spectral regions, and to selectively investigate the environment-induced decoherence. Similarly to pump-probe, this technique is non-linear (resting on multiple laser pulses to be shed on the sample) and has fs resolution. However, being staged in a two-dimensional frequency space (excitation and detection frequencies), it can additionally discriminate between homogeneous and inhomogeneous line-shapes, and allows one to separate the spectrum of a mixture into those of individual components. As a typical example, two-dimensional spectroscopy was applied to visualize long-lasting coherent dynamics of electronic states in bacterial photosynthetic reaction centers \cite{Lee_2007a,Engel_2007a} and carotenoids \cite{Ostroumov_2013a}.

% photoluminescence and electroluminescence
{\bf Watching emitted photons.} In all cases in which donor and acceptor are weakly- or non-interacting in the ground state, CT is still possible through interaction in the excited states. In this situation photoluminescence, both stationary and time-resolved, is a primary tool in the identification of radical species. Quenching of the donor photoluminescence yield and reduction of its lifetime when the donor is mixed with the acceptor, are both signs that one or more CT channels are competing with the radiative recombination in the donor. Time-resolved photoluminescence techniques have nowadays reached a resolution of the order of 100 fs \cite{Kersting_1993a}.

Photoluminescence allows us to differentiate between energy and charge transfer, since the acceptor emission, which appears in the case of energy transfer, is absent in presence of CT. In particular, emission bands at longer wavelengths than emission from the separated donor or acceptor (appearing when the two are mixed) are a signature of CT excitons \cite{Loi_2007a}. Photoluminescence has also been used in its spatially resolved version and combined with microscopy in several flavors to map the interfacial areas where excitons are generated \cite{Coffey_2007a,Wong_2012a,Manca_2013a,Fuchs_2016a,Jakowetz_2017a}.

Charge transfer rates can be estimated by means of time-resolved photoluminescence spectroscopy, i.e. by measuring the luminescence decay rate. However, this information is limited to the incoherent regime, due to the low time resolution.

However, luminescence quenching alone cannot distinguish between a charge separation process yielding free carriers (step (4) in table \ref{tab:steps}), and, for instance, Dexter electron transfer yielding radical pairs in a bound state (as in step (3) of table \ref{tab:steps}). When applied to organic solar cells, electroluminescence can also be used to estimate the CT binding energy: when injected charge carriers recombine at the interface between the donor and the acceptor, a linear correlation between the CT exciton emission and the open-circuit voltage of the solar cell can be established \cite{Tvingstedt_2009a,Veldman_2009a}. Also establishing a link between photoluminescence quenching and the yield of photogenerated free charges can be problematic in situations for which competing recombination paths coexist \cite{Ohkita_2008a}.

% Time-resolved photoemission
{\bf Watching outgoing electrons.} So far, all the techniques we reviewed consisted in measuring the amount of incoming or outgoing electromagnetic radiation after photoexcitation of the system. Similarly to some of the techniques examined in the previous paragraphs, time-resolved photoelectron spectroscopy has order fs resolution, and is sensitive to both vibrational and electronic dynamics. However, ultrafast photoelectron spectroscopy probes one-electron removal spectra, i.e. charged excitations. It is therefore ideal to probe radiationless phenomena and investigate lifetimes and relaxation pathways of excited electronic states in molecules as well as ultrafast non-adiabatic processes \cite{Stolow_2004a}. As such, it has been employed to estimate the coherent size, localization and relaxation times of CT excitons at donor-acceptor interfaces \cite{Wang_2017a} and the kinetics of injection in dye-sensitized interfaces \cite{Borgwardt_2016a}.

% Photocurrent and pump-push
Photocurrent measurements are traditionally used to assess the performance of photoconversion and optoelectronic devices. They are most profitably coupled to ultrafast spectroscopy by employing a suitable sequence of laser pulses. In pump-push-photocurrent spectroscopy, for instance, an intermediate infrared pulse is used to excite (or dump) specific vibrational degrees of freedom. The photocurrent can then be monitored as a function of the delay time between the pump and the push pulses \cite{Bakulin_2012a}. The all-optical analogue of this technique is known as pump-push-probe spectroscopy \cite{Kee_2014a}.

% spin resonance, x-ray diffraction, electron 
On passing, we mention few other powerful techniques based on the use of X-ray coherent sources. Structural modifications and transient photogenerated phase transitions have been highlighted in bulk solids by means of femtosecond X-ray diffraction, which employs an optical pump and X-ray probe, to track ultrafast lattice deformations \cite{Cavalleri_2006a}. Similarly, the structural response in bulk solids can be followed by ultrafast electron crystallography, i.e. employing an optical pump, and an electronic probe \cite{Zewail_2006}. Core-hole spectroscopy may reveal charge transfer dynamics on time scales of the order of the core lifetime (1 fs) and is particularly useful when studying molecular adsorbates on metallic surfaces \cite{Keller_2004,Menzel_2008,Kohntopp_2016}. X-ray pump X-ray probe experiments can excite inner-shell electrons, providing an extremely localized, site-specific probe for excited state dynamics \cite{Picon_2016}.

We finally mention that spin resonance has also been used to reveal the presence of cationic and anionic species \cite{Sariciftci_1992a,Niklas_2015a}.

%%%%%%%%%%%%%%%%%%%%%%%%%%%%%%%%%%%%%%%%%%%%%%%%%%%%%%%%%%%%%%%%%%%%%%%%%%%%%%%%
\section{Ultrafast charge separation in prototypical systems}\label{sec:proto}
%%%%%%%%%%%%%%%%%%%%%%%%%%%%%%%%%%%%%%%%%%%%%%%%%%%%%%%%%%%%%%%%%%%%%%%%%%%%%%%%

Here, we examine the ultrafast photophysical and photochemical properties of a few prototypical organic systems that display ultrafast charge separation. They mostly serve as photovoltaic materials. For a review on photoinduced electron transfer in bonded donor--acceptor systems (not specifically focusing on the ultrafast character) see, for example, Ref. \cite{Wrobel_2011a}. At the current state of the art, these systems cannot be modeled atomistically {\em ab initio}. Two main pathways are thus available to theoreticians: either identify and isolate the most relevant fragments of the systems to which Density-Functional Theory (DFT) or its time-dependent extension (TDDFT) can be applied, or map the supposedly relevant degrees of freedom onto a few-level model, to be solved as accurately as possible.

%%%%%%%%%%%%%%%%%%%%%%%%%%%%%%%%%%%%%%%%%%%%%%%%%%%%%%%%%%%%%%%%%%%%%%%%%%%%%%%%
  \subsection{Supramolecular assemblies}\label{sec:supra}
%%%%%%%%%%%%%%%%%%%%%%%%%%%%%%%%%%%%%%%%%%%%%%%%%%%%%%%%%%%%%%%%%%%%%%%%%%%%%%%%

% Why they are important
Light-harvesting, the primary step of natural photosynthesis, proceeds by collecting sun light at a molecular antenna array and transferring its energy to the reaction center, where it is stored as electrical potential. Natural reaction centers are complex structures, made of a large number of chromophores, which are embedded in a 10-30 nm sized protein scaffold \cite{McDermott_1995a}. 

Supramolecular structures -- typically donor-acceptor dyads, or donor-bridge-acceptor triads -- are an ideal playground for the theoretical and numerical investigation of similar ultrafast charge separation processes. They have smaller dimensions than natural reaction centers, and therefore a small CT exciton radius and limited delocalization extension. They may be considered as models for nano-sized photovoltaic devices, and synthesis protocols exist for testing them as potential light-harvesters, artificial reaction centers, or solar fuels generators \cite{Sherman_2014a}.

% How they are made
Porphyrins are by far the most common building blocks for electron donors and molecular sensitizers in artificial photosynthetic models \cite{Imahori_2004b}. However, other different chormophores, such as diaminoterephthalate, have been proposed also as molecular scaffolds for easier functionalization \cite{Freimuth_2015a,Pittalis_2015a}. In particular, non-bonded dyads made of porphyrins and fullerenes were used to assess constrained DFT and TDDFT accuracy in describing excitation energies \cite{Ghosh_2010a,Cramariuc_2006a,Toivonen_2006a}.

In orgranic photovoltaics, organic chromophores have been linked to several electron acceptors, such as quinones and fullerenes. Fullerene and its functionalized derivatives were soon preferred to quinones, because it was observed that they can considerably accelerate charge separation, while at the same time slowing down charge recombination \cite{Imahori_2001a,Imahori_2004a}.

Fullerenes possess a number of properties that are desirable in this context. In fact, they are soluble in biological membranes. Besides, they show: small solvent and internal reorganization energies upon reduction; radical anions rather insensitive to the solvent dielectric constant (compared e.g. to quinones); a recombination channel yielding triplet excited state, instead of the ground state \cite{Gust_2000a}. In dyads, a CT state can often be reached in less than 60 fs, through coupling in the excited state \cite{Chapman_2011a}.

% Triads
Regarding morphology, triads have an efficiency advantage over dyads, since the recombination of the $(D^{+\bdot}A^{-\bdot})$ state is typically several orders of magnitude slower than in dyads \cite{Kobori_2005a}. This occurs because in triads a second electron transfer from the donor to the bridge unit is competing with the acceptor to donor recombination. This fact mimics, on a smaller scale, what happens in nature: the recombination ratio is substantially reduced when electrons are transferred in a multi-step process. Therefore, supramolecular triads may also be considered as simple artificial surrogates of natural reaction centers.

% Photophysics
The photophysics of supramolecular triads has been extensively studied. In carotenoporphyrin-fullerene (CPC$_{60}$) triads, for example, photoinduced electron transfer leading to the intermediate $CP^{+\bdot}C_{60}^{-\bdot}$ state and to the final $C^{+\bdot}PC_{60}^{-\bdot}$ (see figure \ref{fig:triad_gust}) was observed by means of transient absorption spectroscopy and fluorescence measurements \cite{Liddell_1997a,Kuciauskas_2000a,Bahr_2000a}. Quantum yields, transfer rates and the energetics of the intermediate and final products were measured with ps time resolution, indicating that electron transfer from the carotene to the porphyrin radical cation occurs with an overall yield of 0.95 in a time scale of 125 ps \cite{Kodis_2004a}.

% fs regime
Due to its amenable size, the sub 100 fs dynamics in the CPC$_{60}$ triad can be studied by combining quantum simulations and ultrafast spectroscopy. It was found that the through-bond exciton delocalization between the caroteno-porphyrin and the fullerene is actually coherently driven by the vibrational modes of the molecule. Moreover, a series of computational experiments (see figure \ref{fig:triad_carlo}) suggested that the mobility of individual parts of the molecule, such as the linker unit between the chromophore and the acceptor, can be exploited in order to control the charge separation yield \cite{Rozzi_2013a}.

%%%%%%%%%%%%%%%%%%%%%%%%%%%%%%%%%%%%%%%%%%%%%%%%%%%%%%%%%%%%%%%%%%%%%%%%%%%%%%%%
  \subsection{Dye--sensitized materials}\label{sec:dye}
%%%%%%%%%%%%%%%%%%%%%%%%%%%%%%%%%%%%%%%%%%%%%%%%%%%%%%%%%%%%%%%%%%%%%%%%%%%%%%%%

% Design idea
The idea behind dye--sensitized solar cells consists in the splitting the functions of the charge separation and transport processes into two distinct phases, as opposed to what happens in first generation cells, entirely based on bulk semiconductors junctions \cite{Oregan_1991a}. For a review of the design and operation mode of dye--sensitized electrochemical cells see Ref. \cite{Graetzel_2003}. For a review on modelinig of ultrafast dynamics see Ref. \cite{Prezhdo_2009a}.

In a dye-sensitized material the light-capturing moiety is an organic molecule, the charge-carrier transport phase occurs in a nanostructured semiconductive oxide (typically TiO$_2$), and the contacting phase to the semiconductor is guaranteed by an electrolyte (usually an organic solvent containing the I$^-$/I$_3^-$ redox pair). 

% Function
The basic operation mechanism consists of the following steps: 1) an electron from the photoexcited molecule is transferred to the conduction band of the oxide substrate, leaving the dye in its oxidized state; 2) the electron donor is then restored to its ground state by electron transfer from the electrolyte, once the circuit is closed and the photocurrent can flow. It's clear that in this kind of systems the time scales of charge separation and regeneration determine the efficiency of the device. The charge collection mechanism is controlled by the ratio of the time it takes for the electrons to diffuse through the oxide and their recombination time. 

% Ultrafast kinetics
Focusing on the above step 1), the interfacial electron kinetics is determined by the competition between ultrafast electron injection and excited-state relaxation. Simulations of the electron and hole dynamics at the interface of organic sensitizers with TiO$_2$ nanocrystals show that, after excitation, the electron is injected into the semiconductor on a time scale of 125-175 fs, while the geminate hole injection takes a much longer time \cite{Meng_2010a}. Early measurements of transient absorption and fluorescence up-conversion spectra with femtosecond resolution showed that, when the sensitizer is perylene, reactants-decay and products-raise times for the electron transfer reaction have an identical time constant of 190 fs (in case of perylene as sensitizer), for a wide range of temperatures \cite{Burfeindt_1996a}. Ultrafast electron injection dynamics in the order of 100 fs time scale or faster was confirmed for Ru and other dye-sensitizers \cite{Hannappel_1997a,Tachibana_1996a,Martini_1998a}. Depending on the nature of the materials, injection times can be much slower (up to 100 ps). However, the efficiency of electron injection in dye-sensitized solar cells does not depend on the injection kinetics alone, but rather on the interplay between injection and decay kinetics.

% Factors affecting injection time
It's important to note that, in dye-sensitized semiconductors, electron injection occurs on the same (or faster) time scale than the vibrational energy relaxation within the excited states of the molecules. Therefore, the wave packet motion of the electron along the potential energy surface of the reaction (see figure \ref{fig:marcus}) evolves from a vibrationally excited state of the dye, and not from the ground state \cite{Damrauer_1997a}. The coherent motion of the vibrational wave packet continues even in the product state \cite{Willig_2000a}.

The injection time seems not to be very affected by the energetic alignment between the dye donor and semiconductor acceptor states \cite{Borgwardt_2016a}. On the other hand, it dramatically depends on other factors, such as the electronic coupling between the dye $\pi^*$ orbitals and the oxide semiconductor band: weakening this coupling (for instance, by means of chemical spacers) naturally leads to longer time constants \cite{Spettel_2016a}, and back to a situation in which injection is slower than vibrational relaxation. Similarly, very different time scales are measured for different substrate oxides and in absence/presence of the electrolyte \cite{Antila_2014a}.

Porphyrin dyes constitute another interesting class of systems for applications in dye-sensitized solar cells technology. Recently, these molecules have attracted a lot of interest in the solar cell community, thanks to their tunability (through chemical functionalization) and their panchromatic properties. Porphyrin dyes, like the ones reported in the upper panel of figure~\ref{Fig:Porphyrin}, have a characteristic donor-bridge-acceptor structure, with a strong `vectorial' component that promotes the formation of an intra-chromophore charge separated state right after the photon absorption. In the best case, like for instance for the dye SM315~\cite{mathew2014dye}, the charge separation is such that the photoexcited electron is found in close proximity to the link with the semiconductor (i.e., the TiO$_2$), providing the best conditions for an efficient charge injection (figure~\ref{Fig:Porphyrin}, lower panel). In addition, the density of excited states in this class of porphyrin dyes implies a very broad absorption spectrum that extends from 400 to 800 nm, covering most of the UV-vis solar spectrum (panchromatic effect).

% the sensitizer adsoprtion geoemtry affects the open circuit potential \cite{DeAngelis_2007a}

%%%%%%%%%%%%%%%%%%%%%%%%%%%%%%%%%%%%%%%%%%%%%%%%%%%%%%%%%%%%%%%%%%%%%%%%%%%%%%%%
  \subsection{Organic bulk heterojunctions}\label{sec:bulk}
%%%%%%%%%%%%%%%%%%%%%%%%%%%%%%%%%%%%%%%%%%%%%%%%%%%%%%%%%%%%%%%%%%%%%%%%%%%%%%%%

% Concept
The bulk heterojunctions class includes materials obtained by blending a p-type (hole conducting) and an n-type (electron conducting) organic species \cite{Yu_1995a}: typically, an organic polymer and a fullerene derivative, such as PCBM \cite{Sariciftci_1992a}. After processing and annealing the mixture, the interface of the heterojunction emerges as the complex surface delimiting two well-separated yet inter-penetrating phases. The phases have crystal domains tens of nanometers wide, although the blend as a whole does not show long-range crystalline order. Charge separation occurs at the interface of the two phases.

% Advantages and disadvantages for photovoltaics
This structure solves at least one of the problems in organic photovoltaics, namely the short exciton diffusion length \cite{Street_2016a}. The tight interleaving of the two phases in the mixture has two advantages over a flat bi-layer junction design: 1) it makes the charge-separating interfacial area large, with respect to the film thickness; 2) it keeps the surface between the phases within a distance comparable to the exciton diffusion length of the light absorber. In fact, polymer photoluminescence quenching is usually as high as $95\%$ in these materials. As a drawback, however, the amorphous blend operation as a solar cell is much affected by the nanoscale morphology, because it requires continuous percolating pathways in each phase, for the charges to be collected at the electrodes \cite{Hoppe_2004a}. The details of the morphology also affect electronic and structural properties of these materials, making their systematic investigation challenging \cite{Erb_2005a,Chen_2011a,Liu_2011a,Turner_2011a}.

% Ultrafast generation
The time-resolved photo-induced absorption spectra of pristine polymers dramatically change when the polymer is mixed to an electron acceptor such as PCBM. This fact was taken as indicative of ultrafast photoinduced electron transfer \cite{Brabec_2001a}. Since this observation, many groups have reported ultrafast exciton quenching on a scale of 100 fs in polymer:PCBM heterojunctions \cite{Hwang_2007a,Cook_2009a,Piris_2009a,Howard_2010a,Kaake_2012a,Gelinas_2014a,Falke_2014a}. The production of free charges, however, is believed to be a slower process, occurring on the scale of few to hundreds of ps \cite{Hwang_2008a}. However, it was also reported that the Coulomb binding attraction can be overcome on a time scale as short as 40 fs \cite{Gelinas_2014a}, suggesting that delocalization may be a crucial factor affecting the system dynamics. The detailed interpretation of ultrafast charge separation and the identification of the factors affecting it are still subject of debate; they will be discussed in section \ref{sec:cur}.

% Contribution to the photocurrent
Unfortunately, it seems difficult to find a connection between ultrafast exciton generation and device efficiency. In fact it was shown that in several blends with internal quantum efficiency $>90\%$ still $40\%$ of the excitons are dissociated on a diffusion-limited time scale (1-100 ps) \cite{Kaake_2013a}. In organic bulk heterojunctions the charge transfer time is orders of magnitude faster than the recombination channels. This leads a fraction of about 60\% of the bound electron-hole pairs to yield free charges under short-circuit conditions \cite{Mihailetchi_2004a}. Interestingly, by comparing photoluminescence to photocurrent generated in MEH-PPV:C$_{60}$ blends, it was observed that below-gap radiation contributed negligibly to the photocurrent in comparison to above-gap excitation ~\cite{Drori_2010a}. This clearly indicates that direct generation of CT excitons, although possible with very small cross sections, is not an important charge generation mechanism in these materials. Therefore, ultrafast charge separation in these devices is influenced by several factors including: frontier level alignment (which provides a driving force for charge separation), kinetic competition between generation and recombination pathways, and complex morphology (affecting exciton diffusion length and charge collection) \cite{Zhong_2015a,Jakowetz_2016a}.

%%%%%%%%%%%%%%%%%%%%%%%%%%%%%%%%%%%%%%%%%%%%%%%%%%%%%%%%%%%%%%%%%%%%%%%%%%%%%%%%
\section{Current views on the fundamental charge separation dynamics}\label{sec:cur}
%%%%%%%%%%%%%%%%%%%%%%%%%%%%%%%%%%%%%%%%%%%%%%%%%%%%%%%%%%%%%%%%%%%%%%%%%%%%%%%%

% Intro
In this section we focus on the paradigmatic case of bulk heterojunctions. For these systems, although it is generally accepted that the ultrafast charge dynamics (of the order of 100 fs) is relevant for charge separation, there is no general agreement on the nature of the physical mechanisms governing the dynamics.

As briefly anticipated in sections \ref{sec:intro} and \ref{sec:bulk}, given the large exciton binding energy and the ultrafast character of the charge dynamics in organic systems, one can ask what is the origin of the energy required to to produce the free charge carriers. Several factors have been proposed to solve this contradiction. In the present and following sections, we consider as possible explanations the excess of initial kinetic energy, the delocalization of the CT state, quantum coherence, and vibronic coupling.

%%%%%%%%%%%%%%%%%%%%%%%%%%%%%%%%%%%%%%%%%%%%%%%%%%%%%%%%%%%%%%%%%%%%%%%%%%%%%%%%
\subsection{Excess energy and hot states}\label{sec:hot}
%%%%%%%%%%%%%%%%%%%%%%%%%%%%%%%%%%%%%%%%%%%%%%%%%%%%%%%%%%%%%%%%%%%%%%%%%%%%%%%%

% Hot carriers and excess energy
Some authors found a tight correlation between the free energy difference of the charge separation ($\Delta G_{CS}$) and the yield of long-lived charge generation, and have proposed that the contradiction between energetics and dynamics is easily solved by invoking the excess energy of the incoming photon \cite{Ohkita_2008a}. If the charge-separated state is not produced by dissociation of the lowest-lying CT exciton, as suggested in Table \ref{tab:steps}, but rather by the dissociation of a state having higher energy than the charge-separated one, then the CT state can naturally dissociate in very short time (see figure \ref{fig:views}). In this scenario, an excited ``hot" CT state is the actual precursor of free charge carriers, while the relaxed CT exciton is in fact a trap state, i.e. a photocurrent loss channel. This idea was further supported by femtosecond non-linear optical spectroscopy \cite{Bakulin_2012a,Jailaubekov_2012a,Grancini_2012a,Schulze_2014a}. According to this view, the excess of incoming energy is channeled through strongly coupled electronic and nuclear degrees of freedom, whose nature has not been entirely clarified. It is however clear that this scenario leads to a charge separation mechanism in which thermalization of the hot state and its dissociation to a charge separated state compete, since both occur on the same time scale of hundreds of fs \cite{Pensack_2010a}. Once the charge transfer exciton is allowed to relax to its ground state, it will basically recombine according to a mono-molecular geminate process.

% Against excess energy
Other authors, instead, closely follow the order sketched in Table \ref{tab:steps}, and report that mobile carriers are generated via a multi-step process, in which an intermediate CT bound state, generated within about 100 fs, subsequently gives birth to free charges \cite{Hwang_2008a}. Supporting this view is the observation that, for a wide class of organic heterojunctions, the internal quantum efficiency is essentially independent of excess energy of the charge transfer exciton  \cite{Lee_2010a,Vandewal_2014a}. This interpretation is however in stark contrast, for example, with the observations in Ref.  \cite{Grancini_2012a}, where a wavelength and laser-intensity dependence of the internal quantum efficiency is reported. This matter is made intricate by the difficulty of accurately measuring internal quantum efficiency in stacked structures. Comments and replies on this point can be followed in REf. \cite{Grancini_2013a}. Also, lower efficiencies in systems with lower values of $\Delta G^0_{CT}$ were reported, contradicting the prediction of the hot exciton framework \cite{Howard_2010a}. The ``cold" excitons scenario can be described within a Onsager-Braun model \cite{Onsager_1938a,Braun_1984a}, and implies CT relaxation on a 100 fs time scale, followed by a much slower CT dissociation.

Another source of excess energy is the band offset between the donor LUMO and the acceptor HOMO. Although the correlation between the yield of bound and free charges is observed to be strongly dependent on such energy difference \cite{Clarke_2008a,Bakulin_2013a,Dimitrov_2014a}, it was observed that charges are generated as efficiently in blends with a small charge separation free energy difference (0.2 eV in PCPDTBT:PCBM) as in P3HT:PCBM (for which $\Delta G_{CS}= 0.9$ eV) \cite{Clarke_2009a}.

% Hot excitons and delocalization
Hot excitons have a larger degree of delocalization with respect to the relaxed ones, so that exciton delocalization must enter the debate. This point was investigated in Ref.  \cite{Bakulin_2012a} by interposing an infrared pulse between the usual pump and probe in a transient absorption technique. This extra ``push" pulse interacts with the localized CT state, and results in an increase of the photocurrent with respect to the case when no push was used, within 200 fs. The effect was attributed to an increased delocalization of the bound charge pairs. The beneficial effect of delocalization in charge separation has also been confirmed by kinetic Monte Carlo \cite{Deibel_2009a}, stochastic Schr\"odinger equation \cite{Abramavicius_2016a} and TDDFT simulations \cite{Nan_2015a}.

% Delocalization per se
At this point it is legitimate to ask how delocalization of CT states affects ultrafast dynamics in organics, independently on whether the direct precursor of the charge-separated state is hot or cold. As a matter of fact, exciton delocalization is also affected by other factors, besides the kinetic energy initially provided by incoming photons: among them, primarily the Coulomb electrostatic screening provided by electric dipoles at the interface, and the dielectric screening in each material, especially on the polymer donor side. The combined effect of screening and delocalization could therefore provide an explanation for charge dissociation that does not rest on hot excitons \cite{Nenashev_2011a},  \cite{Bassler_2015a}. Hot excitons, in turn, would be a privileged channel ensuring higher delocalization by exploiting the incoming photon energy.

% Towards coherence
Recently, coherence and uncertainty were thrown into the discussion of charge separation in photovoltaics \cite{Kaake_2013a}, following the idea that a universal quantum feature could be the source of a more effective delocalization. This idea raised criticisms about the extent to which such fundamental concepts can be straightforwardly applied to describe charge separation \cite{Mukamel_2013a}. Even though the central focus remains excited state delocalization in disordered materials \cite{Kaake_2013b}, still, the link between quantum coherence and delocalization has raised further debate in the field, which leads us to the following Section.

% Energy transfer
%At the end of this discussion we mention, that, following a quite independent line of thought, other authors have suggested the possibility that the primary process upon photoexcitation in a P3HT:PCBM thin film is in fact ultrafast {\em energy} transfer from donor to acceptor while charge separation is actually due to hole transfer from acceptor to donor on a much longer time scale (100 ps) \cite{Kandada_2013a}.

%%%%%%%%%%%%%%%%%%%%%%%%%%%%%%%%%%%%%%%%%%%%%%%%%%%%%%%%%%%%%%%%%%%%%%%%%%%%%%%%
  \subsection{Quantum coherence and dissipation to the environment}\label{sec:coh}
%%%%%%%%%%%%%%%%%%%%%%%%%%%%%%%%%%%%%%%%%%%%%%%%%%%%%%%%%%%%%%%%%%%%%%%%%%%%%%%%

We will examine more in depth the different meanings of ``quantum coherence" and how to model them in section \ref{sec:quantcoh}. Here we focus on how the fundamental quantum nature of the microscopic electronic and nuclear states affects the ultrafast dynamics of charge separation in natural and artificial systems.

% Is coherence useful?
The idea that a coherent quantum dynamics is exploited by nature in order to optimize the efficiency of energy transfer in photosynthetic complexes is intuitively appealing. A possible, though not conclusive, argument in favour of such idea is related to random walks. In fact, the mean squared displacement of the initial excitation in a one-dimensional system is proportional to the number of steps $N$ in the case of a {\em classical} diffusive motion and to $N^2$ in the quantum case \cite{Kempe_2003a}.
If and how we might exploit this property to design better photovoltaic or solar fuel production devices is a very debated and lively topic of discussions in the scientific community \cite{Romero_2014a,Park_2015a}. While we were preparing this Topical Review, several reviews appeared specifically on this subject \cite{Bredas_2016a,Romero_2017a,Scholes_2017a}.

The recent development of suitable spectroscopical techniques (see section \ref{sec:intro}) has led to the direct observation of quantum beatings on the same time scale as the one of energy transfer, and confirmed that the protein environment is key to allow the excitation to move coherently in space \cite{Lee_2007a,Engel_2007a,Collini_2010a,Panitchayangkoon_2011a}. The observed coherences are surprisingly long-lived also at room temperatures, and they resist the effect of disorder \cite{Hildner_2013a}. Interestingly, vibrational coherences, which are often of the same order as electronic ones in the ground state, rely upon the excitonic interaction to reveal themselves \cite{Plenio_2013a}. It is natural to think that the protein cage protecting the light-harvesting complexes modulates the electron–phonon couplings and, through the motion of charged residues, also changes the local dielectric environment promoting coherent excitation transfer. Coherent intra-chain electronic energy transfer was also observed in conjugated polymer samples with different chain conformations \cite{Collini_2009a}, exporting this ideas to the field of photovoltaic charge separation. This could also be investigated by a recently reported technique employing scanning tunnel microscope to map the spatial distribution of the excitonic coupling in molecular arrays with sub-nm resolution \cite{Zhang_2016}.

% Coherence and charge-transfer
From a theoretical point of view, it's currently unfeasible to explicitly solve the full quantum equations of motion for organic macromolecules embedded in their complex environment. However, the crossover between hopping (F\"orster) and coherent propagation mechanisms, and, more generally, the role of quantum coherence, have been intensively explored using parametrized model Hamiltonians \cite{Leegwater_1996a,Ghosh_2010a,Abramavicius_2016a}. Demonstrating the genuine quantum nature of coherence might in some case be not straightforward. Some authors have in fact argued that quantum coherence in the electronic excitation transport can be mirrored to classical coherence in dipole-dipole oscillators \cite{Briggs_2011a,Miller_2012a}. 

% Observation of spatially coherent energy and charge transfer
The study of spatial coherence in the excited states has also become an active field of research. The energy energy transfer over long distances occurring in photosynthetic complexes is apparently made more efficient by the coherent delocalization of the excitations. Delocalized CT states were for instance observed in the light-harvesting antenna of photosystem II \cite{Ahn_2008a}. 
In fact, it has been shown that exciton transfer, charge transfer, and charge separation in multi-chromophore donor-acceptor systems can be approached within a single model framework in an approximate way, leading to the conclusion that efficient charge separation depends on an optimum balance between charge delocalization and energy dissipation \cite{Kocherzhenko_2015a}.
Delocalization is strictly related to spatial coherence (see the following section), and can in fact be viewed as a consequence of the linear superposition of many electronic states, each localized on a different chromophore. Exciton localization is affected by the interplay between the electronic coupling of the different sites, the disorder, and the electron-phonon interactions. Thermal fluctuations of the environment on one side are the main source of dephasing affecting therefore the charge localization. On the other side, though, they have been found to be crucial for the optimization of the energy transport process \cite{Mohseni_2008a,Chin_2013a}. In this respect, energy transmission is thus different from that of quantum information. While the quantum information processing typically requires a fully coherent dynamics, efficient energy transmission seems to rely on a complex interplay of coherent and noisy processes \cite{Caruso_2010a}.

In conclusion, quantum coherence might play a role in charge-separation processes, both because it can produce delocalized states and because it can qualitatively modify the dynamics of exciton transport. The ultimate identification of the degrees of freedom responsible for this enhancement, and the issue if the relevant coherences are electronic, vibrational, or vibronic, remain open questions. We will return on this point in section \ref{sec:quantcoh}.

%%%%%%%%%%%%%%%%%%%%%%%%%%%%%%%%%%%%%%%%%%%%%%%%%%%%%%%%%%%%%%%%%%%%%%%%%%%%%%%%
  \subsection{The role of nuclear motion}\label{sec:nuc}
%%%%%%%%%%%%%%%%%%%%%%%%%%%%%%%%%%%%%%%%%%%%%%%%%%%%%%%%%%%%%%%%%%%%%%%%%%%%%%%%

In the above discussion on quantum coherence , we introduced the function of molecular vibrations (or phonons) both as a channel of internal relaxation and as a set of degrees of freedom that, coupled to the electronic ones, can drive the electron and energy transfer dynamics, possibly exploiting quantum coherence. The importance of the coupling of excitons to specific vibrational modes goes beyond the level of the static reorganization energy as it appears in Marcus' theory, and must be accurately included in any quantum mechanical or semi-classical description of the charge (and energy) transfer dynamics (see section \ref{sec:QD}). Signs of such dynamical effects appear in all the prototypical systems examined in section \ref{sec:proto}.

% El-ph coupling in biological complexes
The coupling of electronic excitations to molecular vibrations in biological complexes is often easy to spot, since one or few vibrational modes may exhaustively describe the reaction coordinates along which the charge or energy transfer processes occur. Similarly, in primary events of vision a specific vibrational mode drives the photoisomerization reaction of the retinal chromophore (even though this process occurs often through a conical intersection) \cite{Wang_1994a,Polli_2010a}. The exciton-phonon coupling is well described by theory at different levels of abstraction. For example, the distribution of the vibrational energy can be initially studied by estimating the spectral density of the system, while subsequent normal-modes analysis can better evaluate the impact of each mode on the charge motion \cite{Eisenmayer_2012a,Tiwari_2013a,Jing_2012a}. Details about theoretical modeling methods can be found in Section \ref{sec:QD}. This coupling is proposed to be the source of long-lived oscillations observed in time-resolved spectroscopies in several systems (see section \ref{sec:exp}). In addition, some authors stress the ultimate quantum nature of vibration-assisted exciton transport and claim that it is achieved via non-classical fluctuations of collective pigment motions \cite{OReilly_2014a}.

% In supramolecular bonded systems
In section \ref{sec:supra} we mentioned that in covalently bonded donor-bridge-acceptor structures the CT is controlled by specific vibrations \cite{Rozzi_2013a}. This information provides important clues for designing charge-separating devices, since it has been experimentally shown that, employing infrared pulses, it is actually possible to control the bridge vibrational modes, and hence modulate the dynamics and the yield of the electron transfer process \cite{Delor_2014a,Delor_2015a}. A similar effect has been observed in hydrogen-bonded complexes \cite{Lin_2009a} and rationalized in terms of partial disruption of the O-H bonds \cite{Eisenmayer_2014a}.

% No nuclear displacement with no electronic coherence
In small organic molecules, ultrafast double hole transfer was also observed after core ionization. Interestingly, in this case the electronic states involved are separated by a large energy gap and no electronic coherence is developed between them. In particular, the hole transfer is driven by specific and tiny nuclear displacements (of the order of a few tenths of an \AA) along a non-Born-Oppenheimer path \cite{Li_2015a}.

% In dye-sensitized and polymers
In dye-sensitized TiO$_2$, non-adiabatic electron injection from the dye into the semiconductor turns out to be mediated by vibration of interfacial Ti-O bonds \cite{Jiao_2011a}. In conjugated polymers (often used as donors and/or acceptors in organic photovoltaics) the dynamical coupling of electronic and nuclear degrees of freedom is evident already in the ground state, as prominent vibronic replica of the main exciton feature appear in the linear absorption spectra \cite{Street_2014a}. Intra-molecular vibrational modes can be activated in the excited state in less than 25 fs, and their coherence may last as long as 1 ps \cite{Lanzani_2003a}. In polymers, ultrafast quantum dynamics can be monitored using 2D spectroscopy experiments, which show evidence of coherent charge oscillations between excitons and polaron pairs \cite{Song_2014a,DeSio_2016a}.

% In bulk heterojunctions
The effect of electron-nuclear coupling has also been intensively studied in non-bonded systems, such as bulk heterojunctions. In these systems, the stretching mode of the conjugated acceptor backbone is tightly coupled to the electronic transitions, although Frenkel to CT exciton transitions usually involve slower ring torsional modes. In some cases, ``bridge" states are invoked as CT mediators \cite{Kanai_2007a,Smith_2015a}. In polymer-polymer heterojunctions, the ultrafast exciton decay is rather robust when proceeding through suitable bridging states \cite{Tamura_2008a}. Overall, several state-of-the-art quantum dynamics models agree with the picture that the interplay of electrostatic confinement with the coherent coupling of electrons to vibrational modes controls the CT efficiency across organic interfaces \cite{Bera_2015a,Tamura_2011a}. Electron-phonon interaction is found to enhance the coupling between a donor-localized excitonic state and the CT state, and provides an oscillatory dynamics superimposed to the charge accumulation on the acceptor \cite{Falke_2014a}. These findings altogether promote the idea of vibration-assisted charge separation. Also in this case it was experimentally demonstrated that the excitation of specific modes in the infra-red band leads to photocurrent enhancement \cite{Bakulin_2015a}.

% Final words
This whole body of evidences leads us to deeply rethink the traditional view of charge dynamics based on the Born-Oppenheimer approximation. Although electronic excitations occur on time scales at which the nuclei can indeed be considered fixed, the energy redistribution through vibronic coupling can well take place on time intervals faster than a typical vibrational period. Once created, a vibronic wave packet is immediately able to move along the potential energy surface and mediate the exchange of energy among the exciton and the molecular vibrations.

In the adiabatic picture (i.e. in the regime of strong electronic coupling between the donor and the acceptor), the nuclear reconfiguration drives the reaction through transition states that gradually change their localization from the donor to the acceptor, while the electrons remain on the same nuclear potential energy surface. This requires a good amount of coupling between different vibrational modes, but allows for little energy exchange between excitons and vibrations. However, when the dynamics along nuclear and electronic coordinates is strongly coupled, the adiabatic approximation ceases to be valid, and the time evolution of the CT state to the charge-separated one occurs at a speed comparable to that of the energy dissipation.

%%%%%%%%%%%%%%%%%%%%%%%%%%%%%%%%%%%%%%%%%%%%%%%%%%%%%%%%%%%%%%%%%%%%%%%%%%%%%%%%
\section{Quantum coherence}\label{sec:quantcoh}
%%%%%%%%%%%%%%%%%%%%%%%%%%%%%%%%%%%%%%%%%%%%%%%%%%%%%%%%%%%%%%%%%%%%%%%%%%%%%%%%

As mentioned in section \ref{sec:coh}, the concepts of coherence and entanglement have entered the debate concerning the processes of charge and energy transfer. In fact, quite a number of works specifically addresses the possible role of non-classical features in enhancing the efficiency of natural and artificial light-harvesting systems. Here we introduce theoretical tools that allow, on the one hand, to quantify such features, and, on the other hand, to establish formal relations between coherence, entanglement, and delocalization.

%%%%%%%%%%%%%%%%%%%%%%%%%%%%%%%%%%%%%%%%%%%%%%%%%%%%%%%%%%%%%%%%%%%%%%%%%%%%%%%%
  \subsection{Definitions}\label{sec:cohdef}
%%%%%%%%%%%%%%%%%%%%%%%%%%%%%%%%%%%%%%%%%%%%%%%%%%%%%%%%%%%%%%%%%%%%%%%%%%%%%%%%

Within the framework of classical physics, coherence and interference have been extensively investigated as properties of waves. In quantum mechanics, all systems can in principle exhibit wavelike properties. Such possibility results from the linear superposition principle, according to which, a linear combination of two or more quantum states $|i\rangle$ is still a legitimate quantum state, $ | \psi \rangle = \sum_i c_i |i\rangle $. As a result, the expectation value in $|\psi\rangle$ of an observable differs in general from the weighted average (with weights $p_i=|c_i|^2$) of its expectation values in the component states $|i\rangle$. Such difference represents a manifestation of quantum interference. Quantum states can also be summed incoherently, giving rise to mixed states, such as $ \hat\rho_{inc} = \sum_i p_i | i \rangle\langle i | $. Here, the phase relation between the component states $|i\rangle$ is completely undefined, and no interference shows up in the statistics of the physical observables. Coherence between any two component states $|i\rangle$ and  $|j\rangle$ is formally reflected by off-diagonal terms in the density matrix, $ \rho_{ij} = \langle i | \hat\rho | j \rangle $, also referred to as {\it coherences}. The diagonal elements of the density matrix, $\rho_{ii}$, represent instead the occupation probabilities of the basis states, and are referred to as {\it populations}.

As appears from the above, quantum coherence is a basis-dependent concept. In fact, one of the relevant distinctions between coherences observed in light-harvesting systems is based on the reference basis and on the involved degrees of freedom. In particular, the photoinduced ultrafast dynamics that eventually leads to charge separation involves both the electronic and the nuclear (vibrational) degrees of freedom. One can thus choose a reference basis formed by product states $|i\rangle = |e_i\rangle \otimes |v_i\rangle$, where $e_i$ and $v_i$ define the electronic and vibrational states, respectively. Given the system density operator $\hat\rho$, with matrix elements $\rho_{ij}$, we observe that
\begin{itemize}
  \item $\rho_{ij}$ corresponds to a purely {\it electronic} coherence if the reference states $|i\rangle$ and $|j\rangle$ differ from each other in the electronic degrees of freedom ($e_i \neq e_j$), but not in the nuclear ones ($v_i=v_j$);
  \item more specifically, $\rho_{ij}$ is a {\it spatial} electronic coherence if $|i\rangle$ and $|j\rangle$ differ by the spatial localization of one or more electrons. Coherent delocalization of the exciton states is related to the presence of this kind of coherences;
  \item $\rho_{ij}$ is a {\it vibrational} coherence if $|i\rangle$ and $|j\rangle$ share the same electronic state ($e_i = e_j$), but differ in the state of the vibrational modes ($v_i \neq v_j$). Vibrational coherences result from the creation of phonon wavepackets, resulting from vertical transitions induced by ultrafast laser pulses;
  \item $\rho_{ij}$ is a {\it vibronic} coherence if $|i\rangle$ and $|j\rangle$ differ both in their electronic and in their vibrational components. The presence of these coherences demonstrates a nontrivial interplay between the electronic and the nuclear degrees of freedom, and can result in quantum correlations between the two. It thus calls for theoretical approaches that go beyond the adiabatic approximation and a semiclassical description of the nuclear dynamics.
\end{itemize}
A further distinction between quantum coherences is based on whether $\hat\rho$ represents an eigenstate of the system or a non-stationary state. In particular, one can distinguish between static and dynamic coherences.
\begin{itemize}
  \item {\it Static} coherences are the ones present in an eigenstate $| E_k \rangle$ of the system Hamiltonian $\hat H$, namely $\rho_{ij} = \langle i| E_k \rangle\langle E_k | j\rangle $. The presence of static coherences thus reflects the fact that different states of the reference basis $\{ |i\rangle \}$ contribute to the composition of the eigenstate in question. Static spatial coherences might enhance the efficieny of the energy transfer in light-harvesting systems. Unfortunately, they are not directly accessible in experiments such as optical spectroscopies.
  \item In the case of {\it dynamic} coherences, instead, one refers to a non-stationary state $\hat\rho (t)$. In particular, the coherences between different eigenstates can be generated by an external drive, such as the pump pulse(s) in transient spectroscopies. Their presence thus reflects the way in which the system is manipulated, rather than its intrinsic properties, nor it provides informations concerning the character of the involved eigenstates $i\rangle$ and $|j\rangle$. Dynamic coherences are experimentally accessible, and are in fact responsible for the oscillating features observed, e.g., in 2D spectroscopies (see below). They tend to vanish within a characteristic time scale due to environment-induced decoherence.
\end{itemize}

The term coherence can also be referred to the time evolution of the state $\hat\rho$ as a whole, rather than to one of its off-diagonal matrix elements. In particular, the time evolution $\hat\rho (t_0) \longrightarrow \hat\rho(t>t_0)$ is fully coherent if it's unitary, and thus reversible, and can be described by the time-dependent Schr\"odinger equation. The degree of coherence of the system dynamics is a basis-independent quantity, and does not have a straightforward relation to the off-diagonal elements of the density matrix in some specific reference basis. However, in the representative case of a free evolution of the system, departures from a coherent evolution typically consist in the damping of the off-diagonal terms in the eigenstate basis (dephasing), followed, on a longer time scale, by changes in the populations $\rho_{ii}$ (relaxation and incoherent excitation). As mentioned above, it has been argued that the (approximately) coherent character of the system dynamics might enhance the efficiency of the energy transfer with respect to an incoherent (and thus ``classical'') one, thanks to quantum interference between the different pathways. In the following, unless differently specified, by {\it coherence} we mean any off-diagonal matrix element of $\hat\rho$. 

%%%%%%%%%%%%%%%%%%%%%%%%%%%%%%%%%%%%%%%%%%%%%%%%%%%%%%%%%%%%%%%%%%%%%%%%%%%%%%%%
    \subsection{General coherence quantifiers}\label{sec:cohquant}
%%%%%%%%%%%%%%%%%%%%%%%%%%%%%%%%%%%%%%%%%%%%%%%%%%%%%%%%%%%%%%%%%%%%%%%%%%%%%%%%

Quantum coherence represents a resource in different areas of quantum technology. This has fueled an intense effort to controllably generate linear superpositions in diverse physical systems, protect them from the decoherence processes, and develop tools that allow a quantitative characterization of coherence. Such characterization has been provided within the so-called {\it resource theories}, where the coherence resource can be acquired at a certain cost and is consumed to perform tasks of interest by means of constrained (incoherent) operations \cite{Streltsov_2017}. In the following, we report a few coherence quantifiers that, besides having a clear physical meaning, can directly relate the degree of coherence of a given state $\hat\rho$ to its off-diagonal matrix elements, the coherences $\rho_{ij}$. Under suitable approximations, they can also be expressed in terms of quantifiers of electron delocalization and entanglement. All these quantifiers share some general properties, defined within an axiomatic approach to coherence \cite{Baumgratz_2014a}: they vanish if and only if the state in question is incoherent ($\hat\rho_{inc}=\sum_i p_i | i\rangle\langle i|$); besides, their value does not increase by mixing density matrices together (convexity) nor, on average, by means of selective measurements (monotonicity). 

Applications of the coherence quantifiers to the characterization of the excitation energy transfer are still limited \cite{Sarovar_2010a,Vatasescu_2015a,Vatasescu_2016a}. In order to compute these quantifiers, one needs to derive the system state $\hat\rho$. While ab initio approaches (such as those based on DFT) typically don't provide the full quantum state of the system, they can be used to validate approaches based on model Hamiltonians, from which the quantum state of the relevant degrees of freedom can be derived. Besides, we show below that, under suitable assumptions, coherence quantifiers can be expressed as functions of the populations alone, and these are more likely to be expressed in terms of quantities accessible to DFT-based approaches. Estimating the coherence quantifiers directly from experimentally accessible quantities is also challenging. The tool of choice for investigating coherent effects, namely 2D spectroscopy, ideally gives access to individual pathways (see below), each resulting from the convolution of many time evolutions of the system state, resulting from different laser pulse sequences. The concept (and thus the quantifiers) of coherence refers instead to a given quantum state, which is necessarily defined at a given time $t$ of a particular time evolution. However, 2D spectroscopy can indeed allow an indirect estimate of the coherence quantifiers, for it provides information on what linear superpositions can be generated in the system and how resilient these are to environment-induced decoherence.

%%%%%%%%%%%%%%%%%%%%%%%%%%%%%%%%%%%%%%%%%%%%%%%%%%%%%%%%%%%%%%%%%%%%%%%%%%%%%%%%
  \subsubsection{Relative entropy}
%%%%%%%%%%%%%%%%%%%%%%%%%%%%%%%%%%%%%%%%%%%%%%%%%%%%%%%%%%%%%%%%%%%%%%%%%%%%%%%%

In the case of a distance-based quantifier $C_D$, the amount of coherence in a state $\hat\rho$ is reduced to the measurement of the minimum distance between $\hat\rho$ and an incoherent state $\hat\rho_{inc}$, as given by some distance measure $D$ between quantum states. The first quantifier of quantum coherence we report results from the choice of the relative entropy \cite{Nielsen_2010a} as a measure of the distance between two quantum states $\hat\rho$ and $\hat\sigma$. The expression of the relative entropy reads:
\begin{equation} 
  D_S(\hat\rho , \hat\sigma) = 
  S(\hat\rho \Vert \hat\sigma) = {\rm Tr} (\hat\rho \log \hat\rho) -  {\rm Tr} (\hat\rho \log \hat\sigma) , 
\end{equation}
where $ S = - {\rm Tr} (\hat\rho \log \hat\rho) $ is the von Neumann entropy. The relative entropy is a non-negative quantity, which vanishes if and only if $\hat\rho=\hat\sigma$. It can be shown that the incoherent state $\hat\rho_{inc}$ that minimizes $S(\hat\rho \Vert \hat\rho_{inc})$ coincides with the diagonal part of $\hat\rho$, namely
\begin{equation}
  \hat\rho_{diag} \equiv \Delta (\hat\rho) =
  \sum_i |i \rangle\langle i | \hat\rho | i \rangle\langle i| = 
  \sum_i \rho_{ii} |i \rangle\langle i| ,
\end{equation}
where $\Delta$ is the dephasing (super)operator.
The resulting expression of the coherence quantifier, known as {\it relative entropy of coherence}, reads \cite{Baumgratz_2014a}
  \begin{equation}
  C_{S} (\hat\rho) = S(\hat\rho_{diag}) - S(\hat\rho) .
\end{equation}
The quantity $ C_{S} $ has a clear physical meaning, namely that the degree of coherence of $\hat\rho$ corresponds to the amount of disorder introduced in the state by fully dephasing it in the reference basis. The value of $ C_{S} $ ranges from 0 (for arbitrary mixtures $\hat\rho_{inc}$ of the states $|i\rangle$) to $\log d$ (for linear superpositions such as $|\phi\rangle = \frac{1}{\sqrt{d}}\sum_{i=1}^d |i\rangle$, where $d$ is the dimension of the Hilbert space). In the case of a pure state $\hat\rho = |\psi\rangle\langle\psi|$, one has that $S(\hat\rho)=0$. The coherence of $\hat\rho$ can thus be expressed as a function of the diagonal matrix elements $\rho_{ii}$ alone, and is an increasing function of the dispersion of the population amongst the states of the reference basis
\begin{equation}\label{eqFT08}
  C_{S} (\hat\rho) = - \sum_i \rho_{ii} \log \rho_{ii} .
\end{equation}

%%%%%%%%%%%%%%%%%%%%%%%%%%%%%%%%%%%%%%%%%%%%%%%%%%%%%%%%%%%%%%%%%%%%%%%%%%%%%%%%
  \subsubsection{$l_1$-norm}
%%%%%%%%%%%%%%%%%%%%%%%%%%%%%%%%%%%%%%%%%%%%%%%%%%%%%%%%%%%%%%%%%%%%%%%%%%%%%%%%

The second quantifier of quantum coherence we report is based on a different distance between quantum states, namely that defined by the $l_1$ matrix norm
\begin{equation}
  D_{l_1}(\hat\rho , \hat\sigma) = 
  \Vert \hat\rho - \hat\sigma \Vert_{l_1} = \sum_{i,j} | \rho_{ij} - \sigma_{ij} | . 
\end{equation}
Also in this case, the incoherent state $\hat\rho_{inc}$ that is closer to $\hat\rho$ is $\hat\rho_{diag} = \Delta (\hat\rho)$. The resulting expression of the quantifier, referred to as the $l_1$-{\it norm of coherence}, is given by \cite{Baumgratz_2014a}
\begin{equation}\label{eqFT05}
  C_{l_1} (\hat\rho) = \sum_{i \neq j} |\rho_{ij}| .
\end{equation} 
The above expression formalizes the intuition that the degree of coherence in a given state depends on the number and amplitude of the off-diagonal terms in the density matrix.  
Its value ranges from 0, in the case of incoherent states $\hat\rho_{inc}$, to $d-1$ for states such as $ | \phi \rangle = \frac{1}{\sqrt{d}} \sum_{i=1}^d |i\rangle $. 

%%%%%%%%%%%%%%%%%%%%%%%%%%%%%%%%%%%%%%%%%%%%%%%%%%%%%%%%%%%%%%%%%%%%%%%%%%%%%%%%
  \subsubsection{Fidelity}
%%%%%%%%%%%%%%%%%%%%%%%%%%%%%%%%%%%%%%%%%%%%%%%%%%%%%%%%%%%%%%%%%%%%%%%%%%%%%%%%

Along the same lines, other quantifiers of coherence can be obtained by starting from different distances between quantum states. The {\it fidelity of coherence}, for example, is based on the use of fidelity ($F$) for defining the distance \cite{Baumgratz_2014a}: 
\begin{equation}
  D_F(\hat\rho, \hat\sigma) = 1 - F = 1 - {\rm Tr} \sqrt{\hat\rho^{1/2} \hat\sigma \hat\rho^{1/2}} .
\end{equation}
In the case of a pure state $\hat\rho = |\psi\rangle\langle\psi|$, the corresponding coherence quantifier, $C_F$, takes the simple form
\begin{equation}\label{eqFT03}
  C_{F} (\hat\rho) = 1 - \sqrt{\sum_{i} \rho_{ii}^2} .
\end{equation} 
Therefore, just as the relative entropy of coherence (\ref{eqFT08}), the fidelity of coherence in a pure state increases with the dispersion of the populations corresponding to the basis states $|i\rangle$.

%%%%%%%%%%%%%%%%%%%%%%%%%%%%%%%%%%%%%%%%%%%%%%%%%%%%%%%%%%%%%%%%%%%%%%%%%%%%%%%%
  \subsubsection{Wigner-Yanase-Dyson skew information}
%%%%%%%%%%%%%%%%%%%%%%%%%%%%%%%%%%%%%%%%%%%%%%%%%%%%%%%%%%%%%%%%%%%%%%%%%%%%%%%%

Following a different approach, the coherence of a state $\hat\rho$ can be directly related to its wavelike character. This in turn represents a source of quantum fluctuations in the value of a given observable $ \hat K = \sum_i k_i |i \rangle\langle i| $, whose eigenstates define the reference basis. In fact, in the case of pure states, quantum fluctuations are the only source of uncertainty in the measurement outcome, whereas mixing represents an additional source of uncertainty in the case of mixed states. The above arguments qualitatively justify the use of the {\em Wigner--Yanase--Dyson skew information} $\mathcal{I} (\hat\rho,\hat K)$ as a coherence quantifier \cite{Girolami_2014a}. The skew information $\mathcal{I}$, which is meant to single out the uncertainties resulting from quantum fluctuations, is given by:
\begin{equation}
  \mathcal{I} (\hat\rho , \hat K) = - \frac{1}{2} {\rm Tr} \left\{ [\sqrt{\hat\rho},\hat K]^2 \right\} .
\end{equation}
Amongst other properties, the above quantity vanishes for density operators that commute with $\hat K$, for which only state mixing contributes to the uncertainty in the measurement outcome. On the other hand, the skew information coincides with the variance of $\hat K$ for pure states $\hat\rho=|\psi\rangle\langle\psi|$,
\begin{equation}
  \mathcal{I} (\hat\rho , \hat K) = 
  {\rm Tr} \left( \hat\rho \hat K^2 \right) - \left[ {\rm Tr} (\hat\rho \hat K) \right]^2 
  = {\rm Var} (\hat K),
\end{equation}
while it represents a lower bound for Var$(K)$ in the case of arbitrary density operators $\hat\rho$.

%%%%%%%%%%%%%%%%%%%%%%%%%%%%%%%%%%%%%%%%%%%%%%%%%%%%%%%%%%%%%%%%%%%%%%%%%%%%%%%%
  \subsection{Quantum coherence and delocalization}\label{sec:cohloc}
%%%%%%%%%%%%%%%%%%%%%%%%%%%%%%%%%%%%%%%%%%%%%%%%%%%%%%%%%%%%%%%%%%%%%%%%%%%%%%%%

In a less general setting, one can establish a direct connection between quantum coherence and other physical properties, such as delocalization or entanglement \cite{Smyth_2012a}. In a number of photosynthetic light-harvesting complexes, for example, the system of interest is formed by a number of well-defined subsystems (chromophores), which provide a local basis for the description of the electronic excitations. In fact, $|i\rangle$ can be identified with the state where the excitation is localized in the $i$-th subsystem. Provided that the coupling between the subsystems is small compared to the single-site energy gaps, the state of a singly excited system can be expanded in the basis of such localized excitations, and each subsystem can be regarded as an effective two-level system. The degree of delocalization, either in the Hamiltonian eigenstates or in the time-dependent density matrix, can thus be quantified by the {\it inverse participation ratio} \cite{Thouless_1974a}
\begin{equation}\label{eqFT02}
  {\rm IPR} ( \hat\rho ) = \sum_i \rho_{ii}^2 .
\end{equation}
The more the state is delocalized, i.e. distributed amongst the different subsystems, the smaller the value of IPR is. In fact, the inverse participation ratio varies from $1/d$ for fully delocalized states to $1$ for completely localized ones. If the state $\hat\rho$ is pure, the delocalization as quantified by IPR implies quantum coherence in the local basis and {\em vice versa}. In fact, by combining (\ref{eqFT03}) and (\ref{eqFT02}), one can write the fidelity of coherence $C_F$ in terms of the inverse participation ratio as
\begin{equation}
  C_F (\hat\rho) = 1-[{\rm IPR}(\hat\rho)]^{1/2}.
\end{equation}
Such coherent delocalization can be related to the linear superposition principle, and is thus a quantum effect. In general, however, such connection cannot be established, and one might have a fully incoherent and yet maximally delocalized state, such as $\hat\rho_{inc} = (1/d) \sum_i |i \rangle\langle i|$. Such incoherent delocalization expresses a classical uncertainty concerning what site $i$ the excitation is localized on, which might result from an ensemble average or from a dephasing process in the localized basis.

A more general connection between delocalization and coherence can be established by means of the {\it coherence length}, which is defined as \cite{Meier_1997a}
\begin{equation}
  CL (\hat\rho) = \frac{\left(\sum_{i,j} |\rho_{ij}|\right)^2}{d\sum_{i,j} |\rho_{ij}|^2} .
\end{equation}
The coherence length varies from $1/d$ for states localized on a single site, to $d$ for fully and coherently delocalized states, while it takes the intermediate value of 1 for fully and incoherently delocalized states. It thus turns a qualitative difference between the coherent or incoherent character of the delocalization, into a quantitative one. It might be instructive to express a coherence quantifier, and specifically $C_{l_1}$, in terms of the coherence length
\begin{equation}
  C_{l_1} (\hat\rho) = \left[ d \,\, CL(\hat\rho) \, {\rm Tr} (\hat\rho^2) \right]^{1/2} - 1 .
\end{equation} 
In the above expression, the relation between quantum coherence, coherence length and state purity (quantified by $1/d \le {\rm Tr}(\hat\rho^2) \le 1$) clearly emerges: large values of the coherence length can correspond either to large or to small values of $C_{l_1}$, depending on whether $\hat\rho$ is a pure or a highly mixed state.

%%%%%%%%%%%%%%%%%%%%%%%%%%%%%%%%%%%%%%%%%%%%%%%%%%%%%%%%%%%%%%%%%%%%%%%%%%%%%%%%
  \subsection{Quantum coherence and entanglement}\label{sec:cohent}
%%%%%%%%%%%%%%%%%%%%%%%%%%%%%%%%%%%%%%%%%%%%%%%%%%%%%%%%%%%%%%%%%%%%%%%%%%%%%%%%

Entanglement arguably represents the most characteristic feature of quantum systems. In the last decades, it has been extensively investigated, also in view of its potential role in quantum technologies \cite{Amico_2008a,Horodecki_2009a}. Entanglement and coherence are distinct and yet related concepts. At a basic level, one might argue that entanglement results from the application of the superposition principle to composite quantum systems. More specifically, entanglement-based quantifiers of coherence have been defined, based on the intuition that quantum coherence in a given system $S_1$ is required in order to generate entanglement between $S_1$ and a second system $S_2$ through incoherent operations \cite{Streltsov_2015a}. In a less general framework, one can derive direct relations between coherence and entanglement quantifiers or witnesses \cite{Sarovar_2010a,Smyth_2012a}. In particular, we consider the above mentioned single-excitation manifold, spanned by the states $|i\rangle$, each one corresponding to the excitation being localized in the $i$-th two-level subsystem. The reduced density matrix of two subsystems $i$ and $j$, expressed in the basis of the Fock states $|n_i,n_j\rangle$ and specifically $|0,0\rangle$,$|0,1\rangle$,$|1,0\rangle$,$|1,1\rangle$ takes the form
\begin{eqnarray}
  \hat\rho^{(2)} (i,j) & = & 
  \left(
  \begin{array}{cccc}
    \sum_{k\neq i,j} \rho_{kk} & 0         & 0         & 0 \nonumber\\
    0                          & \rho_{ii} & \rho_{ij} & 0 \nonumber\\
    0                          & \rho_{ji} & \rho_{jj} & 0 \nonumber\\
    0                          & 0         & 0         & 0 
  \end{array}
  \right) .
\end{eqnarray} 
This form of the reduced density matrix reflects the constraints resulting from the presence in the system of a single excitation. As a result, the occupation probability corresponding to a simultaneous excitation of the two subsystem ($ | 1,1\rangle $) vanishes identically, and no coherence is allowed between states corresponding to different values of $n_i+n_j$. The degree of entanglement between two-level subsystems can be quantified by the concurrence $\mathcal{C}$ \cite{Wootters_1998a}, which varies from 0 (for separable states) to 1. Concurrence is typically used in qubit systems, with few exceptions trying to incorporate it within an {\em ab initio} many-body framework \cite{Pittalis_2015b}. In this particular case, $\mathcal{C}$ can be written as a simple function of the coherence between the two states corresponding to $n_i+n_j=1$, i.e. $ | 0,1 \rangle $ and $ |1,0 \rangle $ \cite{Sarovar_2010a}: 
\begin{equation}\label{eqFT04}
  \mathcal{C} \left[ \hat\rho^{(2)} (i,j) \right] = 2 |\rho_{ij}| .
\end{equation}
A direct relation between quantum coherence and entanglement can be derived by combining the expression of the $l_1$-norm of coherence (\ref{eqFT05}) and that of the concurrence (\ref{eqFT04}), which leads to
\begin{equation}
  C_{l_1}(\hat\rho) = \sum_{i < j} \mathcal{C} \left[ \hat\rho^{(2)} (i,j) \right].
\end{equation}
Thus, the overall amount of entanglement between pairs of subsystems, quantified by the sum of the concurrences on the right-hand side of the above equation, corresponds exactly to the amount of coherence in the system state, as quantified by $C_{l_1}$. The above relation does not depend on the specific definition of the subsystems $i$, provided that the above mentioned structure of the Hilbert space is preserved, as well as the constraint on the overall number of excitations. Analogous relations between purity, entanglement and IPR can be derived \cite{Smyth_2012a}

%%%%%%%%%%%%%%%%%%%%%%%%%%%%%%%%%%%%%%%%%%%%%%%%%%%%%%%%%%%%%%%%%%%%%%%%%%%%%%%%
\section{Theoretical spectroscopies}\label{sec:thspect}
%%%%%%%%%%%%%%%%%%%%%%%%%%%%%%%%%%%%%%%%%%%%%%%%%%%%%%%%%%%%%%%%%%%%%%%%%%%%%%%%

As already mentioned, charge and energy transfer, charge separation, and, in general, the processes of harvesting and converting solar light take place in large and complex structures, whose full description in terms of the Schr\"odinger equation would largely exceed the available computing capabilities. One way to overcome such limitation is to explicitly include only a limited number of relevant degrees of freedom in the simulation of the quantum dynamics. This leads us to the content of this section, in which  we review how to describe open quantum systems. These approaches are the means of choice for the simulation of nonlinear spectroscopies, which in turn represent the most powerful experimental probe of excited state dynamics. See also section \ref{sec:exp} for a review of the diverse experimental techniques, and sections \ref{sec:proto}-\ref{sec:cur} for a number of applications to the study of representative systems.

%%%%%%%%%%%%%%%%%%%%%%%%%%%%%%%%%%%%%%%%%%%%%%%%%%%%%%%%%%%%%%%%%%%%%%%%%%%%%%%%
  \subsection{Dynamics of open quantum system}\label{sec:openqs}
%%%%%%%%%%%%%%%%%%%%%%%%%%%%%%%%%%%%%%%%%%%%%%%%%%%%%%%%%%%%%%%%%%%%%%%%%%%%%%%%

The dynamics of an isolated quantum system is described by the Schr\"odinger equation. In reality, due to the unavoidable coupling to a physical environment, there exists no such thing as an isolated quantum system. On the other hand, even in complex systems, it is often possible to identify a set of relevant degrees of freedom, which are experimentally accessible and weakly coupled to the remaining ones in the system. These degrees of freedom define the {\it reduced system} $\mathcal{S}$, while the remaining ones constitute the {\it environment} $\mathcal{E}$. The time evolution of the reduced density operator, $\hat\rho_\mathcal{S}$, is in principle obtained by evolving the complete state (through the unitary time-evolution operator $\hat U_\mathcal{SE}$) and then tracing over the environment,
\begin{equation} \label{eqFT09}
  \hat\rho_\mathcal{S} (t) = {\rm Tr}_\mathcal{E} \left\{ \hat U_\mathcal{SE} (t)\, \left[ \hat\rho_\mathcal{S} (0) \otimes \hat\rho_\mathcal{E} (0) \right] \, \hat U_\mathcal{SE}^\dagger (t) \right\} .
\end{equation} 
In practice, it is desirable to express the reduced dynamics in a closed form, where the environment enters implicitly through an approximate description of its effect on the time evolution of $\hat\rho_\mathcal{S}$. This can be done through two equivalent but formally different approaches, namely quantum master equations and linear maps, which are briefly discussed hereafter \cite{Breuer_2007a}. In the following, we refer to $\mathcal{S}$ simply as the {\it system} and drop the subscript $\mathcal{S}$ from $\hat\rho_\mathcal{S}$, unless this is required to avoid ambiguities. 

\subsubsection{Quantum dynamical maps}

The dynamics of the system resulting from Eq. (\ref{eqFT09}) can be expressed, without any explicit reference to the environment, in terms of a {\it quantum dynamical map} \cite{Breuer_2007a}
\begin{equation}
  \Lambda_{0,t} : \hat\rho (0) \longrightarrow \hat\rho (t) ,
\end{equation}
which is a linear superoperator, hermiticity- and trace-preserving, and completely positive. These properties guarantee that the dynamics preserves the formal properties of the density operator throughout its time evolution. In particular, positivity implies that the eigenvalues of the density operator, representing occupation probabilities, always remain non-negative (as it is the case in the initial state). Complete positivity is an even stronger condition, corresponding to the positivity of the maps $\mathcal{I}_n \otimes \Lambda_{0,t}$, with $\mathcal{I}_n$ the identity operator in an Hilbert space of the environment of arbitrary dimension $n$. This condition guarantees that, also in the case where $\mathcal{S}$ is initially entangled with an $n$-dimensional environment, the application of the map $\Lambda_{0,t}$ to the state of $\mathcal{S}$ (and of the identity operator to that of $\mathcal{E}$) leads to a physical state $\rho_\mathcal{SE}$ of the overall system. 

A {\it quantum dynamical semigroup} is a family of quantum dynamical maps, indexed by a continuous time parameter $t$, that satisfy the semigroup property 
\begin{equation}
  \Lambda_t \Lambda_s = \Lambda_{s+t}. 
\end{equation}
This property can be used to define a Markovian dynamics, along the lines of what is done for classical systems in the homogeneous case. 

If the linear map is invertible, i.e. if a map $\Lambda_{0,t}^{-1}$ exists for all values of $t$, one can also introduce the notion of {\it divisibility} and define a map that couples states corresponding to any two finite times $s$ and $t>s$, namely 
\begin{equation}
  \Lambda_{s,t}=\Lambda_{0,t}\Lambda_{0,s}^{-1}. 
\end{equation}
We note that (complete) positivity of the inverse map $\Lambda_{0,t}^{-1}$, and thus of $\Lambda_{s,t}$, does not follow from that of $\Lambda_{t}$. However, if the dynamical map $\Lambda_{s,t}$ is (completely) positive, then the map $\Lambda_t$ is said to be ({\it completely}) {\it positive divisible}. Completely positive divisibility represents another possible criterion to define Markovianity of a quantum dynamics, in analogy to that of classical processes. As discussed below, the above properties of the dynamical map $\Lambda_t$ can be related to those of the corresponding master equation. 

The definition of quantum Markovianity can also be derived from a connection between memory effects and the information flow between system and environment. In particular, a Markovian dynamics corresponds to an unidirectional flow of information, resulting in a loss of distinguishability between the system states \cite{Breuer_2009a,Laine_2010a}. This is in turn reflected in a monotonic decrease of the distance between any two system states, as quantified by some suitable metrics, such as the trace distance \cite{Nielsen_2010a} 
\begin{equation} 
  D (\hat\rho, \hat\sigma) = \frac{1}{2} \Vert \hat\rho - \hat\sigma \Vert = \sum_k |\mu_k|, 
\end{equation}
where $\mu_k$ are the eigenvalues of $\hat\rho-\hat\sigma$. Interestingly, one can show that the application of a quantum dynamical map decreases the trace distance or, in the case of a unitary dynamics, leaves it unaffected \cite{Ruskai_1994a}
\begin{equation}
  D[\hat\rho(0), \hat\sigma(0)] \ge D \{ \Lambda_t [\hat\rho(0)] , \Lambda_t [\hat\sigma(0)] \} .
\end{equation} 
This property, however, does not prevent the trace distance at time $t$ from being larger that that at time $s$, with $ 0 < s < t $ (as mentioned above, $\Lambda_{s,t}$ might be undefined, or it can exist but without being a quantum dynamical map). If this is the case for some pair of initial states $\hat\rho(0)$ and $\hat\sigma(0)$, then the dynamics described by $\Lambda_t$ is defined non-Markovian. In view of the above definitions, in the case of a divisible map $\Lambda_{0,t}$ the notion of Markovianity is equivalent to that of completely positive divisibility, whereas maps that are positive, but not completely positive divisible describe a non-Markovian dynamics.

\subsubsection{Quantum master equation}

The quantum master equation is a first-order differential equation, which expresses the time derivative of the system state at time $t$ as a function of $\hat\rho$ at time $t$ and, eventually, at earlier times $s<t$. In the simplest case, the master equation is local in time, and the generator $\mathcal{L}$ is time independent
\begin{equation}
  \frac{d}{dt} \hat\rho(t) = (\mathcal{L}_0 + \mathcal{L}_1) \hat\rho(t) 
  = (i/\hbar) [\hat\rho(t), \hat H] + \mathcal{L}_1 \hat\rho(t),
\end{equation}
where $\mathcal{L}_1$ accounts for the coupling to the environment, while $\mathcal{L}_0 $ describes the contribution of the system Hamiltonian $\hat H$. The above master equation reduces to the Liouville-von Neumann equation in the special case $\mathcal{L}_1=0$. 

Relations between the Markovian character of a quantum master equation and that of the corresponding quantum dynamical map have been established. In particular, it can be shown that, under general mathematical conditions \cite{Breuer_2007a}, a quantum dynamical semigroup can be expressed in the exponential form $\Lambda_t = e^{\mathcal{L}t}$, where the generator $\mathcal{L}$ is a Lindblad superoperator. Such time evolution of the density operator can also be expressed in terms of a master equation in the Lindblad form, where
\begin{equation}\label{eqFT07}
  \mathcal{L} \hat\rho = \frac{i}{\hbar} [\hat\rho, \hat H] + \frac{1}{2} \sum_{k=1}^{d^2-1} \Gamma_k \left( 2\hat L_k \hat\rho \hat L_k^\dagger - \hat L_k^\dagger \hat L_k \hat\rho - \hat\rho \hat L_k^\dagger \hat L_k \right).
\end{equation}
Here $d$ is the dimension of the Hilbert space, $\Gamma_k \ge 0$, and the Lindblad operators $\hat L_k$ define an orthonormal operator basis. One can tentatively assign to each of the terms entering the above superoperator a phenomenological meaning. In particular, the $k$-th Lindblad operator can correspond to a given transformation $\hat L_k$ induced by the environment on the system state, at a rate $\Gamma_k$. However, one should also keep in mind that, for a given master equation, the Hamiltonian $\hat H$ and the Lindblad operators are defined up to a set of transformations, under which the superoperator $\mathcal{L}$ is invariant. In fact, the Hamiltonian $\hat H$ appearing in the above equation generally does not coincide with the free Hamiltonian of the reduced system. 

The master equation can be derived either phenomenologically or from the microscopic Hamiltonian of the overall system \cite{Breuer_2007a,Cohen_2004a}. The latter derivation generally does not lead to a master equation in the Lindblad form. In order for this to happen, a number of approximations must be introduced, which we briefly mention hereafter. First of all, one needs to introduce the Born approximation \cite{Breuer_2007a}, which holds if the system-environment coupling is weak, compared to those that enter the free Hamiltonian of the system $\mathcal{S}$. This allows one to expand the complete equation of motion to second order in the system-environment interaction Hamiltonian, to assume that the state of the environment is unaffected by the system, and to write the overall density operator $\hat\rho_{\mathcal{SE}}(t)$ in the factorized form $\hat\rho_{\mathcal{S}}(t) \otimes \hat\rho_{\mathcal{E}}(0)$. As to the Markov approximation, this is valid if the time scale over which the correlation functions of the environment vanish is much smaller than the time scale characterizing the dynamics of the reduced system. The above approximations lead to the so-called Redfield master equation, which is local in time, but  does not necessarily generate a quantum dynamical map, and therefore might lead to an unphysical density operator. The Lindblad master equation is obtained by additionally introducing the secular approximation. 

More generally, if the dynamical map $\Lambda_t$ is invertible, the dynamics can always be expressed in terms of a time-local master equation $d\hat\rho (t) / dt = \mathcal{L} (t) \hat\rho(t) $, where $\mathcal{L}$ represents a generalized Lindblad superoperator, with time-dependent parameters $\Gamma_k$, and Lindblad operators $\hat L_k$. Besides, if the map is a semigroup, all rates and operators have to be time-independent, and a necessary and sufficient condition for complete positivity is represented by the condition that all the rates be nonnegative \cite{Gorini_1976a}. The generalized Lindblad master equation mentioned above is derived through a projection operator technique \cite{Shibata_1977a,Chaturvedi_1979a}. In this approach, the partial trace on the environment degrees of freedom is associated to a projector operator $\mathcal{P}$ in the space of the system environment degrees of freedom. The projected density operator 
$ \mathcal{P} \hat\rho_\mathcal{SE} = {\rm Tr}_\mathcal{E} (\hat\rho_\mathcal{SE}) \otimes \hat\rho_{\mathcal{E}0} = \hat\rho_\mathcal{S} \otimes \hat\rho_{\mathcal{E}0} $ (with $\hat\rho_{\mathcal{E}0}$ a reference state of the environment) is identified with the relevant part of $ \hat\rho_\mathcal{SE} $, for which one seeks a closed equation of motion. The time-convolutioness projection operator technique leads to a master equation local in time, with a time-dependent generator. Such master equation supports an investigation of non-Markovian effects beyond the Born approximation. An alternative projection operator technique leads to the so-called Nakajima--Zwanzig master equation \cite{Nakajima_1958a,Zwanzig_1960a}, which is an integro-differential and time-nonlocal equation, where the derivative of $ \mathcal{P} \hat\rho_\mathcal{SE} $ at time $t$ depends on the past history of such operator through a memory kernel. 

Markovian master equations find their natural application in quantum optics, where the environment is essentially represented by a continuum of modes of the electromagnetic field, whose correlation time is typically shorter than the relevant time scales in the system dynamics. Memory effects might emerge in the case of environments with long correlation times, in the presence of system-environment couplings that are comparable to those within the system, or of a structured environment. Some of these conditions can be met in semiconductor \cite{Khaetskii_2002a} or superconducting \cite{Yoshihara_2006a} qubits, as well as in optomechanical systems \cite{Groblacher_2015a}, molecular spins \cite{Troiani_2008a}, and photosynthetic complexes \cite{Rebentrost_2011a}, just to name a few. Some authors also report that both Markovian and non-Markovian dynamics affect the generation and time evolution of entanglement in biological light-harvesting systems \cite{Caruso_2010a}.

In an open quantum system, the coherences between the system eigenstates tend to be suppressed by the interactions with the environment. Within the framework of quantum-information processing, different strategies have been developed in order to preserve quantum coherence, ranging from active, error-correction approaches \cite{Shor_1995a} to passive, error-avoiding ones \cite{Lidar_2014a}. The ultimate goal of such approaches is to avoid uncontrolled deviations of the system dynamics from the specific unitary evolution that implements the desired kind of quantum information processing. More specifically, one might ask under which conditions some particular coherences can persist in an open quantum system in spite of its coupling to an environment. Indeed, one can show that, for suitable initial conditions, the amount of coherence in a given basis $\{ |i\rangle \}$ can be strictly conserved ({\it freezing of coherence}) if the dephasing process acts in a basis transversal to $\{ |i\rangle \}$ \cite{Bromley_2015a}. Besides, in the presence of a non-Markovian dephasing, a certain degree of coherence between the system eigenstates can persist in the stationary state, a phenomenon known as {\it coherence trapping} \cite{Addis_2014a}. 

%%%%%%%%%%%%%%%%%%%%%%%%%%%%%%%%%%%%%%%%%%%%%%%%%%%%%%%%%%%%%%%%%%%%%%%%%%%%%%%%
  \subsection{Simulation of 2D spectroscopy}\label{sec:nonlinspec}
%%%%%%%%%%%%%%%%%%%%%%%%%%%%%%%%%%%%%%%%%%%%%%%%%%%%%%%%%%%%%%%%%%%%%%%%%%%%%%%%

In transient-absorption spectroscopies, one measures the absorption spectrum of a system of interest that has been previously brought out of its equilibrium state. Typically, the system is excited (pumped) by an ultrashort laser pulse and probed by a successive pulse, with varying time delay between the two. Two-dimensional (2D) spectroscopy represents an extension of such a pump-probe scheme. In fact, it allows to resolve and correlate the excitation and the emission/absorption frequencies. In the time-domain approach, this is achieved by exciting the system with two broad-band femtosecond laser pulses, separated by a variable time delay $t_1$ and followed, with delay $t_2$, by a third, probe pulse. Two-dimensional spectroscopy gives access to nonlinear optical effects, and typically to third-order nonlinearities. In fact, the three electric fields corresponding to the incoming laser pulses induce a polarization in the sample, which results in the emission of a new field in the phase matched direction. Hereafter, we briefly recall the relevant quantities that are accessed in 2D spectroscopy, the relation between such quantities and quantum coherence, and between the simulations of the 2D spectra and the theory of open quantum systems. We refer the reader to the review articles and to the textbooks on 2D spectroscopy for a more detailed discussion of its many aspects \cite{Mukamel_1995a,Hamm_2011a}.

The optical response of the system can be derived from its polarization $P$, which thus represents the key quantity in the following discussion. In particular $P$ is proportional to the energy exchanged by the probe pulse with the system, resulting in negative (absorption) or positive (emission) contributions. Within the dipole approximation, the system polarization is defined as the expectation value of the dipole operator $\hat\mu$, and is a function of the waiting times between the incoming laser pulses
\begin{equation}
  P(t>0) = {\rm Tr} \left[ \hat\rho(t) \hat\mu \right] = P(t_1,t_2,t_3=t).
\end{equation}
Here, $t_1$ is the length of the time interval between the two pump pulses, $t_2$ is the delay of the probe pulse with respect to the second pump pulse, and $t_3$ the time elapsed after the probe pulse. In order to simulate the optical response of the system, one thus needs to compute the time evolution of the density operator $\hat\rho$ from its equilibrium state $\hat\rho (-\infty)$ to some final time $t=T$ of the order of its equilibration time. Within the framework of open quantum systems, such evolution can be computed either exactly (i.e. nonperturbatively) or by means of a perturbative approach. In the following, we refer to the perturbative approach, where the nonlinear contribution to the polarization can be clearly defined. In order to avoid unnecessary complications, we shall assume that the duration of the laser pulses is negligible with respect to the characteristic time scale of the dynamics (semi-impulsive limit) \cite{Mukamel_1995a,Hamm_2011a}.

Within the perturbative approach, the $n$-th order contribution $P^{(n)}$ to the polarization is computed by expanding the equation of motion for the density operator $\hat\rho$ with respect to the light-matter interaction Hamiltonian $ \hat H = - \hat\mu E(t) $, and retaining only those terms that are $n-$th order. In particular, the third-order contribution $P^{(3)}$, which is the dominant nonlinear term in randomly oriented systems with inversion symmetry, is linear in the fields corresponding to each of the three pulses. Being the polarization a linear function of the density operator, $P^{(3)}$ can be identified with the expectation value of $\hat\mu$ obtained from the third-order contribution in the density operator
\begin{equation}\label{eqFT01}
  \hat\rho^{(3)} (t)  = C \,
  \Lambda_{0,t_3} \mathcal{V} \Lambda_{-t_2,0} \mathcal{V} \Lambda_{-t_1-t_2,-t_2} \mathcal{V} \,\hat\rho (\!-\infty) ,
\end{equation}
where
$ C = (i/\hbar)^3 E(0) E(-t_2) E(-t_1-t_2) $, $E(t)$ is the electric field at time $t$ (at the relevant position), $\mathcal{V} \hat\rho \equiv [\hat\mu,\hat\rho]$ is the superoperator that accounts for the light-matter interaction, and  $\Lambda_{t,t+\tau}$ is the quantum dynamical map corresponding to the free evolution of the system. If the duration of the pulses is not negligible with respect to the time scale of the free dynamics, the above equation has to be replaced by a three-dimensional integral with respect to the waiting times. As to the free evolution, the superoperator $\Lambda$ can take different forms, depending on the system Hamiltonian and on the coupling to the environment. In the simplest case, the dynamics is homogeneous with respect to time and the decoherence has a Markovian character. As a result, the time evolution superoperator can be expressed in an exponential form 
$ \Lambda_{t,t+\tau} = \Lambda_\tau = \exp (\mathcal{L}\tau)$, 
with $\mathcal{L}$ a Lindblad superoperator. 

Many features of interest in the 2D spectra result from the coherent contribution to the system time evolution. It might thus be useful to consider the ideal case of a purely Hamiltonian dynamics, where the master equation is reduced to the Liouville--von Neumann equation and the effect of $\Lambda$ on the density operator takes the simple analytical form
\begin{equation}
  \Lambda_\tau [\hat\rho (t)] = \sum_{k,l} \rho_{kl} (t) \, e^{-i\omega_{kl}\tau} 
  | k \rangle\langle l |, 
\end{equation}
where $ \hbar \omega_{kl} = E_k - E_l $, while $|k\rangle$ and $E_k$ are the eigenstates and eigenvalues of the free Hamiltonian $\hat H_0$, respectively. 
The resulting expression of the third-order polarization, for a system initialized in the ground state $|0\rangle$, is given by
\begin{eqnarray}\label{eqFT06}
  P^{(3)} (t) & = & C
  \sum_{\alpha}
  \mathcal{V}_{0           0           , i_{1\alpha} j_{1\alpha}}
  \mathcal{V}_{i_{1\alpha} j_{1\alpha} , i_{2\alpha} j_{2\alpha}}
  \mathcal{V}_{i_{2\alpha} j_{2\alpha} , i_{3\alpha} j_{3\alpha}}
  \nonumber\\
  & \times &
  \langle j_{3\alpha} | \hat \mu |i_{3\alpha} \rangle \,
  e^{-i\omega_{1\alpha}t_1}
  e^{-i\omega_{2\alpha}t_2}
  e^{-i\omega_{3\alpha}t_3}
  .
\end{eqnarray}
Here, the oscillation frequencies are
$ \omega_{k\alpha} \!=\! \omega_{i_{k\alpha}j_{k\alpha}} \!=\! (E_{i_{k\alpha}} \!-\! E_{j_{k\alpha}})/\hbar $, 
while
$\mathcal{V}_{ij,kl}$ gives the amplitude of the dipole-induced transition between
the matrix elements $\rho_{kl}$ and $\rho_{ij}$. Its expression is given by
\begin{equation}
  \mathcal{V}_{ij,kl} 
  = 
  \delta_{lj} \langle i | \hat\mu | k \rangle - \delta_{ik} \langle l | \hat\mu | j \rangle ,
\end{equation}
with $\delta_{ij}$ the Kronecker delta.
As emerges from the above equation, $P^{(3)}(t)$ includes a number of interfering terms, labeled by the index $\alpha$, whose relative amplitudes depend on the above elements $\mathcal{V}_{ij,kl}$. Each of these contributions corresponds to a pathway that leads from the initial state $\rho(-\infty)=|0 \rangle\langle 0|$ to an operator $|i_{3\alpha} \rangle\langle j_{3\alpha}|$ that contributes to the expectation value of $\hat\mu$ (i.e. such that $\langle j_{3\alpha} | \hat\mu | i_{3\alpha} \rangle \neq 0$). Within each pathway, the $k$th laser pulse creates a population or coherence $|i_{k\alpha}\rangle\langle j_{k\alpha}|$ (depending on whether or not $i_{k\alpha}$ coincides with $j_{k\alpha}$), which then oscillates with a frequency $\omega_{k\alpha}$ as a function of the waiting time $t_k$. We stress that these oscillations do not provide information on the character of the eigenstates (e.g. on the spatial coherences that might be present within each of them), but merely witness the presence of a linear superposition of $|i_{3\alpha}\rangle$ and $|j_{3\alpha}\rangle$, amongst others. Besides, each term $\alpha$ correlates the frequencies corresponding to the three waiting times ($\omega_{1\alpha}$, $\omega_{2\alpha}$, and $\omega_{3\alpha}$) and shows that the corresponding coherences evolve one into the other throughout the system dynamics. In particular, in 2D spectroscopy one correlates the frequencies $\omega_1$ and $\omega_3$, obtained by Fourier transforming $P^{(3)}$ with respect to the waiting times $t_1$ and $t_3$, for given values of $t_2$. These quantities are also referred to as the {\it excitation} and {\it detection} frequencies. 

In order to further clarify the above concepts, we briefly discuss here two of the simplest model-systems that include both electronic and vibrational degrees of freedom. The first one is represented by an (electronic) two-level system interacting with a (vibrational) harmonic oscillator, through a Holstein term
\begin{eqnarray}\label{eqFT10}
  \hat H 
  & = 
  \sum_{\chi=g,e} | \chi \rangle\langle \chi | \left[ \epsilon_\chi + \hbar\omega \sqrt{S_\chi} \left( \hat a^\dagger + \hat a \right) \right] + \hbar\omega \hat a^\dagger \hat a
  \nonumber\\
  & = 
  \sum_{\chi=g,e} \sum_{k=0}^\infty 
  | \chi_k \rangle\langle \chi_k | \left[ \epsilon_\chi + \hbar\omega \left( k - S_\chi ,\right) \right]
\end{eqnarray}
where the Hamiltonian eigenstate $|\chi_k\rangle = |\chi , n_\chi \rangle$ is given by the tensor product of an electronic state $\chi = g,e$ and of an eigenstate of the displaced oscillator (corresponding to the annihilation operator $\hat b_\chi = \hat a+\sqrt{S_\chi}$, with $S_\chi$ the Huang-Rhys factor). In spite of its extreme simplicity, such a model can be used to interpret part of the 2D maps obtained with conjugated polymers \cite{Song_2015a} and accounts for the formation of different kinds of coherences, induced by the sequence of three laser pulses (Figure \ref{figFT01}). In particular, the first pulse initially creates purely electronic coherences, formally resulting from the commutator of $\hat\rho (-\infty)$ and of the dipole operator 
$\hat\mu = \mu_{ge} | g \rangle\langle e | + \mu_{eg} | e \rangle\langle g | $. During the waiting time $t_1$, these rapidly evolve into vibronic coherences, as the phonon wavepacket oscillates back and forth within the adiabatic potential energy surface corresponding to the excited state or remains in its initial state, depending on whether the electronic subsystem is in the state $|e\rangle$ or in $|g\rangle$, respectively. Correspondingly, the electronic coherence $\rho_{ge} = {\rm Tr}(\hat\rho |e\rangle\langle g|)$ is periodically suppressed and restored, as the overlap between the reduced vibrational states $\langle e | \hat\rho | e \rangle$ and $\langle g | \hat\rho | g \rangle$ undergoes sequential collapses and revivals. Purely vibrational coherences $ \langle \chi_k | \hat\rho | \chi_l \rangle $ (with $\chi=g,e$) are generated by the second laser pulse (in fact, these are the only coherences supported by the model during the waiting time $t_2$), both in the ground and in the excited-state manifold. Finally, the third laser pulses generates coherences that oscillate with frequencies $ (\epsilon_e-\epsilon_g)/\hbar + k\omega $ as a function of $t_3$ (where, in the $\omega_3$, unlike in $\omega_1$, $k$ can also be negative, as happens in photoinduced absorption).

Additional phenomena can show up in a model system where two optical excitations, $e$ and $f$, are present, and are coupled by an hopping-like term
$\hat H_t = (t/2) (| e \rangle\langle f | + | f \rangle\langle e |)$. 
The Hamiltonian of such a system can be written in the form
\begin{eqnarray}
  \hat H 
  & = 
  \sum_{\chi=g,\lambda,\mu} 
  | \chi_k \rangle\langle \chi_k | \left\{ \epsilon_\chi + \hbar\omega \left[ \hat b_\chi^\dagger \hat b^{}_\chi - (1/2)(S_e+S_f) \right] \right\}
  \nonumber\\
  & + (1/2) (S_e-S_f) (\hat a^\dagger +\hat a) ( | \lambda \rangle\langle \mu | + | \mu \rangle \langle \lambda | ) ,
\end{eqnarray}
where the uncoupled electronic states $|e\rangle$ and $|f\rangle$ are assumed degenerate for simplicity, while
$|\lambda\rangle \equiv (|e\rangle + |f\rangle) / \sqrt{2} $
and
$|\mu    \rangle \equiv (|e\rangle - |f\rangle) / \sqrt{2} $
are the eigenstates of the electronic Hamiltonian. The states $|e\rangle$ and $|f\rangle$ might be identified with the optical excitation being localized in two different, spatially separated subsystem, and $\hat H_t$ with an electronic (or dipole-dipole) coupling between the two subsystems that tends to delocalize the excitation. Irrespective of such an identification, a non-adiabatic term is present in the above Hamiltonian, which possibly represents the minimum model where surface hopping of the phonon wavepacket can take place. In addition, if the electronic gap (here coinciding with $t$) is close to resonance with the phonon energy $\hbar\omega$, the Hamiltonian eigenstates have an hybrid electronic-vibrational character (i.e. vibronic coherences are present in the eigenstates). Such states can provide an efficient pathway for the generation of vibrational coherences, even for small differences between the Huang-Rhys factors of the ground and excited states \cite{Plenio_2013a,Chin_2013a}.

The third-order polarization measured in 2D spectroscopy thus provides information on (differences between) the energy eigenvalues. With respect to a linear response, which provides analogous information, $P^{(3)}(t)$ generally contains additional terms, corresponding to the coherences between excited states. More importantly, multidimensional spectroscopy also allows one to unravel complex optical spectra by correlating excitation and emission frequencies. Such correlations represent an experimental evidence that two or more excitations are dynamically coupled to each other, and thus belong to the same quantum system. In addition, those excitations whose nature has already been clarified can act as labels, and allow one to extract from the 2D spectra a deep insight into the interplay between different degrees of freedom in the system dynamics and in the charge separation process \cite{DeSio_2016a}. 
 
The above illustrative discussion has been carried out by treating the system of interest as an isolated one. In fact, the third-order polarization as expressed in Eq. (\ref{eqFT06}) can be derived from the knowledge of the system Hamiltonian alone. In real systems, however, interactions with the environment cause deviations from such an idealized picture. In the simplest case, this results in an exponential decay of the coherences that are generated by the external driving fields. The resulting third-order polarization can be obtained simply by replacing the real frequencies $\omega_{k\alpha}$ in (\ref{eqFT06}) with complex frequencies $\omega_{k\alpha} - i \gamma_{k\alpha}$, with $\gamma_{k\alpha} > 0$ the decay rate of the coherence in question. In addition, incoherent population transfer between the eigenstates can result from environment-induced relaxation or excitation processes. Finally, couplings between coherences and populations can be induced by the environment, if decoherence is described within a Redfield or (in the presence of degeneracies) Lindblad master equation. All these couplings result in additional pathways, which can complicate the interpretation of the observed 2D spectra. 

The 2D spectroscopy has been identified as the tool of choice for the observation of quantum coherence in a number of physical systems. As mentioned above, the oscillating terms in the third-order polarization $P^{(3)}$ (and, more generally, in the overall polarization $P(t)$) indeed correspond to off-diagonal elements of the density matrix in the basis of the system eigenstates. The observation of oscillations as a function the waiting time $t_k$ thus allow one to detect coherence in the system state after the first $k$ laser pulses. In other terms, it implies that the quantifiers of coherence defined in the previous section take finite values. The extraction of quantitative estimates of the state coherence from the 2D spectra is less straightforward. In fact, besides the remarks reported in the previous section, one should note that the intensity of the field emitted by the system is proportional to the coherence in the density operator that represents the state of the ensemble. However, the constant of proportionality is generally unknown. Besides, peaks corresponding to different coherences often end up in the same spectral region, and might be impossible to resolve, also in view of the inhomogeneous broadening. Experimentally, different pathways can be at least partially discriminated against each other by selecting specific emission directions (phase matching) or by controlling the sum of the pulse phases (phase cycling) \cite{Hamm_2011a}. From a theoretical point of view, the coherence quantifiers can be computed in a straightforward manner, once the quantum dynamics has been solved and the time dependent density operator (or state vector) has been derived.

%%%%%%%%%%%%%%%%%%%%%%%%%%%%%%%%%%%%%%%%%%%%%%%%%%%%%%%%%%%%%%%%%%%%%%%%%%%%%%%%
\section{Approaches to quantum dynamics}\label{sec:QD}
%%%%%%%%%%%%%%%%%%%%%%%%%%%%%%%%%%%%%%%%%%%%%%%%%%%%%%%%%%%%%%%%%%%%%%%%%%%%%%%%
The simulation of ultrafast charge separation processes requires the solution of the entangled electronic and nuclear dynamics.
To this end, several approaches has been developed in the past 10 years.
%A wide variety of methods have been developed for the  simulation of quantum effects in molecular dynamics. 
These can be divided into two main classes depending on whether the nuclei are treated as classical or quantum particles. 
In the former, the nuclear wavepacket is approximated by an ensemble of particles that follow classical trajectories;  quantum corrections are then added (in an approximate way) to deal with nonadiabatic effects.
In the latter, the nuclear wavepacket is described including all quantum effects, such as nonlocality, tunneling and quantum decoherence. %interference between different parts of the packet after scattering at the potential energy surface (PES) crossings.  
In particular the multiconfiguration time-dependent Hartree (MCTDH) algorithm~\cite{meyer90,Beck2000} and \textit{ab initio} multiple spawning (AIMS) dynamics~\cite{martinez1997non} have been used with great success to gain insight into the 
complicated dynamics at conical intersections as well as for the investigation of light-driven reactions in chemistry and biology.
A third class, known as semiclassical methods, adds part of the missing quantum effects to the classical simulations by means of a quantum phase associated to the trajectories; the most common semiclassical methods have been reviewed in recent articles~\cite{topaler98,hack00,Worth2002,meyer90}.

%In order to incorporate the nuclear quantum degrees of freedom into a classical formulation, a number of mixed quantum-classical models have been proposed. 
%In the Born Oppenheimer approximation, the electronic dynamics is evaluated along the classical path of the nuclei, this \textit{ansatz} is often referred to as classical-path approximation. 
Methods that incorporate nuclear quantum degrees of freedom (DoF) into a classical formulation of the dynamics are called
mixed quantum-classical. 
An important limitation common to all these classical path approaches is the absence of a ``back-reaction'' of the classical DoF to the quantum DoF. %dynamics of the quantum DoF.   
One way to overcome this deficiency is to employ Ehrenfest's theorem~\cite{ehrenfest1927} and calculate the effective force on the classical trajectory through a mean potential that is averaged over the quantum DoF~\cite{gerber82, kosloff91, bornrmann96, zhu04}. 
As with most mixed quantum-classical formulations, the resulting mean-field trajectory method employs a quasiclassical approximation to the heavy-particle DoF; that is, the quantum mechanical spread of the initial state %of the subsystem treated classically 
is simulated through a quasiclassical sampling of the corresponding probability distribution. 
This procedure, often called independent trajectory approximation, produces strictly independent trajectories and therefore (in contrast to a rigorous semiclassical description in the sense of the Van Vleck--Gutzwiller formulation~\cite{gutzwiller2000,miller72}) possible quantum-mechanical interferences between individual classical paths cannot be captured.

%In general, a mixed quantum-classical description may be derived by starting with a quantum-mechanically exact formulation  for the complete system followed by a partial classical limit for the heavy-particle DoF. 
%However, this procedure is not unique, since it depends on the particular quantum formulation chosen as well as on the specific way to achieve the classical limit. 
%For example, in the mean field trajectory method the wave-function formulation of quantum mechanics for the total system is adopted and the Ehrenfest classical limit is performed for the heavy-particle DoF. 
%Alternatively, one may consider the Liouville equation of the density operator and perform a classical Wigner limit for the heavy-particle DoF. This leads to the quantum-classical Liouville description, which has recently received considerable attention~\cite{Martens1997,Donoso1998,kapral:8919,horenko2004}. Furthermore, one may start with a path-integral formulation and treat the heavy-particle DoF by the stationary-phase approximation, thus yielding Pechukas' theory~\cite{Pechukas69}. Quite recently, also the hydrodynamic or Bohmian formulation of quantum mechanics has been used as a starting point for a mixed quantum- classical description~\cite{Burant2000, prezhdo2001, Bittner2003, Burghardt2004, wyattbook}. It should be made clear at the outset that the mixed quantum-classical formulations differ greatly, depending on whether the problem is approached via the wave-function, density-operator, path integral, or hydrodynamic formulation of quantum mechanics.

A different way to combine classical and quantum mechanics is given by the ``connection approach'', which was proposed independently by Landau, Zener, and St\"uckelberg and has later been adopted and generalized by many authors~\cite{miller72,herman84,Coker1995, Keppeler03}. In this formulation, nonadiabatic transitions of classical trajectories are described in terms of a connection formula of the semiclassical Wentzel--Kramers--Brillouin~\cite{miller70,heller75a} wavefunctions associated to two or more coupled electronic states. This intuitively appealing picture of trajectories hopping between coupled potential-energy surfaces gave rise to a number of mixed quantum-classical implementations of this idea~\cite{preston71}. 
Recently, a novel coupled-trajectory approach derived from the exact factorization theorem~\cite{abedi_2014} was proposed. In this case, trajectories are propagated on-the-fly along a single time-dependent potential energy surface and are coupled together through the action of the quantum momentum.  This approach named coupled-trajectory mixed quantum-classical (CT-MQC) dynamics  is able to describe (without the need for an \textit{ad hoc} phenomenological model) quantum coherence and decoherence effects that are missing in standard MQC approaches based on classical trajectories.
%In recent years, the term ``surface hopping'' and its underlying concepts have also been used in the stochastic modeling of  deterministic diffusion equations in phase space, such as the quantum-classical Liouville equation.
%In contrast to quasiclassical approximations, semiclassical methods~\cite{miller70,Herman1984,heller91} take into account the phase $exp(iSt /\hbar)$ of a classical trajectory and are therefore capable -- at least in principle -- of describing nuclear quantum  effects. Such processes include nuclear tunneling, interference effects in the wavepacket dynamics, and the conservation of zero-point energy.

In addition, there is a series of alternative methods such as quantum-classical Liouville~\cite{Burant2000, prezhdo2001, Bittner2003, Burghardt2004, wyattbook,curchod2013ontrajectory,tavernelli2013ab} and quantum hydrodynamic or Bohmian formulation~\cite{Burant2000, prezhdo2001, Bittner2003, Burghardt2004, wyattbook,curchod2013ontrajectory,tavernelli2013ab}, which however cannot yet be applied to the description of molecular charge separation processes in fully-atomistic modeling.
In the following sections, we focus on the descriptions of a small subset of quantum dynamics approaches that share the common property of being particularly suited for quantum molecular dynamics simulations at a reasonable computational cost.

%%%%%%%%%%%%%%%%%%%%%%%%%%%%%%%%%%%%%%%%%%%%%%%%%%%%%%%%%%%%%%%%%%%%%%%%%%%%%%%%
  \subsection{Wavepacket-based approaches}\label{sec:wpmeth}
%%%%%%%%%%%%%%%%%%%%%%%%%%%%%%%%%%%%%%%%%%%%%%%%%%%%%%%%%%%%%%%%%%%%%%%%%%%%%%%%
An interesting solution of the electronic and nuclear quantum dynamics is
based on the so-called Multi Configuration Time Dependent Hartree (MCTDH) approach~\cite{meyer90}.
The starting point is the expansion of the total molecular wavefunction according to Born and Huang \textit{Ansatz}
\begin{equation} 
  \Psi(\bsm r, \bsm R, t) =\sum_{i=1} \Phi_{i}(\bsm r; \bsm{R}) \, \Omega_i(\bsm R, t) \, ,
  \label{eq:BH}
\end{equation}
which allows the description of the quantum dynamics as the time-evolution of a nuclear wavepacket evolving on the PESs derived from the solution of the time-independent Schr\"odinger equation for the electrons at fixed  nuclear positions. 
In (\ref{eq:BH}),
$\Psi(\bsm r, \bsm R, t)$ is the molecular wavefunction, $\Phi_{i}(\bsm r; \bsm{R})$ are the static electronic
wavefunctions that depend parametrically on the nuclear coordinates ($\bsm R$), and $\Omega_i(\bsm R, t)$
are the `nuclear' wavefunctions ($\bsm r$ is the collective vector for all electronic coordinates).

In MCTDH the wavefunction \textit{Ansatz} is written in a linear combination of Hartree products
\begin{eqnarray}
  & \Omega_i(\bsm R, t) \rightarrow \Psi_i(Q_1(\bsm R),\dots,Q_f(\bsm R),t) = \nonumber \\
  & = \sum_{j_1=1}^{n_1} \dots \sum_{j_f=1}^{n_f} 
  A^{i}_{j_1 \dots j_f}(t) 
  \prod_{k=1}^f \phi^{i,(k)}_{j_k} (Q_k(\bsm R),t)
  \label{eq:MCTDFwf}
\end{eqnarray}
where $Q_1(\bsm R),\dots,Q_f(\bsm R)$ are the nuclear collective degrees of freedom (i.e. normal modes), $A^{i}_{j_1 \dots j_f}$ are the time-dependent expansion coefficients, and $\phi^{i,(k)}_{j_k}$ are the time-dependent basis function for each nuclear DoF $k$ and each surface $i$. In the most general case, each collective variable $Q_j(\bsm R)$ is therefore a function of all nuclear coordinates, $\bsm R$.

By its very nature, MCTDH requires however the computation of all relevant PESs and corresponding couplings before the actual propagation of the wavepacket. This clearly implies an important computational effort that limits the applicability of this method to a small number of degrees of freedom ($ \leq 10$). This is especially true when the selected degrees of freedom cannot be approximated by model potentials of harmonic or Morse type.

In addition, the determination of the relevant degrees of freedom to include in the dynamics can also become a challenging problem and requires some \textit{a priori} knowledge of the most relevant vibrational modes involved in the process of interest. Finally, MCTDH is not suited for the description of the complex dynamics of nearly chaotic systems, especially when different PES topologies (bound vs. unbound states) need to
be considered. It is worth mentioning that with the advent of the Direct Dynamics variational method (DD-vMCG~\cite{Lasorne15}) this situation may evolve rapidly in the near future~\cite{Vacher2017,Penfold2017}.

%%%%%%%%%%%%%%%%%%%%%%%%%%%%%%%%%%%%%%%%%%%%%%%%%%%%%%%%%%%%%%%%%%%%%%%%%%%%%%%%
  \subsection{Trajectory-based approaches}\label{sec:traj_based_dyn}
%%%%%%%%%%%%%%%%%%%%%%%%%%%%%%%%%%%%%%%%%%%%%%%%%%%%%%%%%%%%%%%%%%%%%%%%%%%%%%%%

%%%%%%%%%%%%%%%%%%%%%%%%%%%%%%%%%%%%%%%%%%%%%%%%%%%%%%%%%%%%%%%%%%%%%%%%%%%%%%%%
    \subsubsection{Mixed quantum-classical trajectory-based approaches.}\label{subsub_MQC}
%%%%%%%%%%%%%%%%%%%%%%%%%%%%%%%%%%%%%%%%%%%%%%%%%%%%%%%%%%%%%%%%%%%%%%%%%%%%%%%%

The first step in the derivation of the equations of motion for the combined electron-nuclear dynamics
is the definition of a suited representation of the total system wavefunction.
Depending on the particular choice of this expansion we can obtain different (approximated)
solutions of the initial molecular Schr\"odinger equation.
In the following we will restrict ourselves to two main representations of the total molecular wavefunction 
that will give rise to two main trajectory-based nonadiabatic molecular dynamics solutions: mean field Ehrenfest dynamics derived from the following {\em Ansatz} for the molecular wavefunction
\begin{equation}
  \Psi(\bsm r, \bsm R,t) = \Phi(\bsm r,t)\Omega(\bsm R,t) e^{\left[ \frac{i}{\hbar}\int^t_{t_0}{E_{el}(t')dt'}\right]}
  \label{Ansatz_eh}
\end{equation}
and surface hopping dynamics based on the Born-Huang expansion (\ref{eq:BH}).
%\begin{equation*}
%\footnotesize{
%\Phi(\bsm r,t)\Omega(\bsm R,t) e^{\left[ \frac{i}{\hbar}\int^t_{t_0}{E_{el}(t')dt'}\right]} \xleftarrow{\text{Ehrenfest }} \Psi(\bsm r, \bsm R,t) \xrightarrow{\text{Born-Huang}} \sum_i^{\infty} \Phi_i(\bsm r; \bsm R)\Omega_i(\bsm R,t)}
%\end{equation*}
%%
In both approaches, the nuclei are described by classical trajectories (one in the mean field case) and therefore the methods belong to the class of  MQC solutions. Comparing the two representations (Born-Huang, (\ref{eq:BH}), vs. Ehrenfest, (\ref{Ansatz_eh})), we observe that while in the second one we have an explicit time-dependence in both nuclear and electronic degrees of freedom, in the Born-Huang expansion the electronic wavefunction, $\Phi_i(\bsm r, \bsm{R})$, only depends implicitly on time through the evolution of the nuclear coordinates, $\bsm{R}(t)$. In fact, in this case the electronic wavefunctions associated to the different states are evaluated from an optimization procedure (self-consistent optimization) at fixed nuclei positions. This implies that non-equilibrium electronic ultrafast processes are better described using the Ehrenfest approach, which incorporates an effective, explicit, electronic dynamics that can also be performed at fixed nuclei positions.
 
It is worth mentioning that other representations of the molecular wavefunction are possible, like the one based on the exact factorization~\cite{abedi2010exact} which leads to a coupled-trajectories MQC approach~\cite{abedi_2014, min_2017} or the one derived from the conditional wavefunction approach~\cite{albareda2014,albareda2015conditional}. Despite their potential, these methods are still in their infancy and therefore they will not be discussed in this review.

%%%%%%%%%%%%%%%%%%%%%%%%%%%%%%%%%%%%%%%%%%%%%%%%%%%%%%%%%%%%%%%%%%%%%%%%%%%%%%%%
    \subsubsection{Ehrenfest dynamics.}
%%%%%%%%%%%%%%%%%%%%%%%%%%%%%%%%%%%%%%%%%%%%%%%%%%%%%%%%%%%%%%%%%%%%%%%%%%%%%%%%

Ehrenfest dynamics (EHD) is derived using the following {\em Ansatz} for the molecular wavefunction
\begin{equation}
  \Psi(\bsm r, \bsm R,t) = \Phi(\bsm r,t)\Omega(\bsm R,t) \exp \left[ \frac{i}{\hbar}\int_{t_0}^{t} d t' \,  E_{el}(t')\right],
  \label{eq:ehr}
\end{equation}
%where $\Phi(\bsm r,t)$ is the electronic and $\Omega(\bsm R,t)$ is the nuclear wavefunction, and
% \begin{equation}
%  E_{el}(t)= \int \int d \bsm r  \,   d \bsm R  \, \Phi^{*}(\bsm r,t)\Omega^{*}(\bsm R,t) \hat{\mathcal{H}}_{el}(\bsm r,\bsm R) \Phi(\bsm r,t)\Omega(\bsm R,t) \, .
% \end{equation}
%Inserting (\ref{eq:ehr}) into (\ref{eq:tdse}) we obtain~\cite{rev}
%\begin{align}
%i\hbar \partial_t \, \Phi(\bsm r,t)&= - \frac{\hbar^2}{2} \sum_i \nabla_i^2 \Phi(\bsm r,t) + \left[ \int{\!\!d\bsm{R}\,\, \Omega^{\ast}(\bsm R,t) \hat{V}(\bsm r,\bsm R) \Omega(\bsm R,t) } \right] \Phi(\bsm r,t) \\
%i\hbar \partial_t \, \Omega(\bsm R,t) &= - \frac{\hbar^2}{2} \sum_{\gamma} M^{-1}_{\gamma} \nabla_{\gamma}^2 \Omega(\bsm R,t) + \left[ \int{\!\! d\bsm{r}\, \,\Phi^{\ast}(\bsm r,t) \hat{\mathcal{H}}_{el}(\bsm r,\bsm R) \Phi(\bsm r,t) } \right] \Omega(\bsm R,t)\,.
%\label{eq:bc06}
%\end{align}
%which in the classical limit for the nuclear degrees of freedom become
which, when inserted in the Schr\"odinger equation for $\Psi(\bsm r, \bsm R,t)$, gives (in the classical $\hbar \rightarrow 0$ limit for the nuclear wavefunction) the following equations of motion for (classical) nuclei  and electrons~\cite{tavernelli06,tavernelli2005molecular}
\begin{equation}
  M_\gamma \ddot{\bsm{R}}_{\gamma}(t)  = - \nabla_{\gamma} \langle \hat{\mathcal{H}}_{el}(\bsm r, \bsm R) \rangle 
  \label{eq:ehcl1}
\end{equation}
\begin{equation}
  i \hbar \partial_t \, \Phi(\bsm r; \bsm R(t), t) = \hat{\mathcal{H}}_{el}(\bsm r; \bsm R(t)) \Phi(\bsm r; \bsm R(t), t).
  \label{eq:ehcl2}
\end{equation}
Here $\langle \hat{\mathcal{H}}_{el}(\bsm r, \bsm R) \rangle=\int{\!\! d\bsm{r}\,\, \Phi^{\ast}(\bsm r,t) \hat{\mathcal{H}}_{el}(\bsm r, \bsm R) \Phi(\bsm r,t) }$, 
\begin{eqnarray}
  & E_{el}(t) = \\ 
  & = \int\!\!\int d \bsm r  \,   d \bsm R  \, \Phi^{*}(\bsm r,t)\Omega^{*}(\bsm R,t) \hat{\mathcal{H}}_{el}(\bsm r,\bsm R) \Phi(\bsm r,t)\Omega(\bsm R,t) \nonumber
\end{eqnarray}
and $ \hat{\mathcal{H}}_{el}(\bsm r, \bsm R)$ is the electronic Hamiltonian in the field generated by the nuclei. 
The non-adiabatic character of this approach can be made clear introducing the representation of the Ehrenfest state as a linear combination of instantaneous adiabatic states functions, $\{\Phi_{I}^{opt}(\bsm r; \bsm{R}(t))\}_{I=1}^{N_s}$, obtained from the solution of the corresponding time-independent Schr\"odinger equation for the nuclear configuration $\bsm{R}(t)$,
\begin{equation}
  \Phi(\bsm r; \bsm R(t), t) =\sum_I^{N_s} c_{I}(t) \, \Phi_{I}^{opt}(\bsm r; \bsm{R}(t)) \, .
  \label{eq:bc17.2}
\end{equation}
Inserting this expansion in (\ref{eq:ehcl1}) leads to the following expression for the nuclear forces
\begin{equation}
  \bsm{F}_K = - \sum_I |c_I|^2 \nabla_K E_I + \sum_{IJ} c^*_I c_J (E_J-E_I) \bsm{d}_{IJ}
  \label{eq:bc17.3}
\end{equation}
where $\bsm{d}_{IJ}=\langle \Phi_{I}^{opt} |\nabla_{\bsm R_K}| \Phi_{J}^{opt} \rangle$ are the non-adiabatic couplings between states $I$ and $J$.

The Ehrenfest equations of motion for the electrons (\ref{eq:ehcl2}) can be easily ``densityfunctionalized" leading to and equation of motion for the Kohn-Sham orbitals~\cite{tavernelli2005molecular,tavernelli06} 
\begin{equation}
  i \hbar \frac{\partial}{\partial t} \phi_k(\bsm r,t) = -\frac{1}{2} \nabla^2 \phi_k(\bsm r,t) +v_s[\rho, \Phi_0](\bsm r,t) \,  ,
\end{equation}
where $\quad k=1, \dots, N_{el}$, 
\begin{eqnarray}
  &v_s[\rho,\Phi_0](\bsm r,t) = \nonumber \\ 
  &=v_{ext}(\bsm r ,t) + v_H(\bsm r,t) + \left. \frac{\delta {E}_{xc}[\rho,\Phi_0](\bsm r)}{\delta \rho(\bsm r)} \right\vert_{\rho(\bsm r)\leftarrow \rho(\bsm r,t)} \, ,
  \label{eq:bc17.0}
\end{eqnarray}
in the adiabatic approximation of the TDDFT kernel, $v_{ext}(\bsm r ,t)=- \sum_{I} \frac{ Z_{I}}{|\bsm r- \bsm R_I |}$ is the external potential, and $v_H(\bsm r,t)$ is the Hartree potential. The dynamics is started from a given initial density (or Kohn-Sham orbitals), which can either correspond to the ground state of the system or to an electronically excited~\cite{tavernelli2005molecular} or ionized~\cite{tavernelli06, gaigeot2010theoretical,lopez2011ultrafast,lopez2013ultrafast} state.

%%%%%%%%%%%%%%%%%%%%%%%%%%%%%%%%%%%%%%%%%%%%%%%%%%%%%%%%%%%%%%%%%%%%%%%%%%%%%%%%
    \subsubsection{Tully's Trajectory Surface Hopping}
%%%%%%%%%%%%%%%%%%%%%%%%%%%%%%%%%%%%%%%%%%%%%%%%%%%%%%%%%%%%%%%%%%%%%%%%%%%%%%%%

Starting from the Born--Huang  expansion of the molecular Hamiltonian~\cite{bornhuang}
\begin{equation}
  \Psi(\bsm r, \bsm R, t)=\sum_{i=1}^{\infty} \Omega_{i}(\bsm R,t) \Phi_{i}(\bsm r;\bsm R)  
  \label{eq:bh}
\end{equation} 
and passing to the classical limit for the nuclei, J. Tully derived the following set of equations for the coefficients ${C}^{[\alpha]}_j(t)$ that are now replacing the quantum nuclear wavefunctions $\Omega_{i}(\bsm R,t) $~\cite{tully90}, 
\begin{eqnarray}
  & i \hbar \dot{C}^{[\alpha]}_j(t) = \nonumber \\
  &\sum_{i=1}^\infty C^{[\alpha]}_i(t) \left(E^{el}_i(\bsm R^{[\alpha]})\delta_{ij} - i \hbar \sum_{\gamma}^{N_{n}}  \bsm d_{ji}^{\gamma}(\bsm R^{[\alpha]})\cdot \dot{\bsm R}_\gamma^{[\alpha]} \right)  \, ,
  \label{eq:tully}
\end{eqnarray}
where  $\bsm{d}_{ji}^{\gamma}(\bsm R) = \int d\bsm r \, \Phi_j^*(\bsm r; \bsm R) \nabla_{\gamma} \Phi_i(\bsm r; \bsm R) $ are the nonadiabatic coupling vectors, NACV. 
%A rigorous formulation of NACVs within linear response TDDFT is given in Refs.~\cite{tavernelli09,Tavernelli10b}.  
Note that in the expansion of (\ref{eq:bh}) the electronic state wavefunctions are depending only implicitly on time through the nuclear coordinates $\bsm R(t)$, which, according to Tully's prescription, evolve `adiabatically' on a given PES $j$ until a \textit{hop}  to a different surface $i$ occurs, with a probability that is computed from the state amplitudes $\{C^{[\alpha]}_i(t)\}$~\cite{tully1991nonadiabatic}. Tully's surface hopping is a multi-trajectory approximation of the exact nuclear dynamics, and it is the ensemble of trajectories (labelled by the superscript $[\alpha]$) that ultimately describes the semiclassical time-evolution of the nuclear wavepacket.

%%%%%%%%%%%%%%%%%%%%%%%%%%%%%%%%%%%%%%%%%%%%%%%%%%%%%%%%%%%%%%%%%%%%%%%%%%%%%%%%
\section{The electronic structure problem}\label{sec:wfmeth}
%%%%%%%%%%%%%%%%%%%%%%%%%%%%%%%%%%%%%%%%%%%%%%%%%%%%%%%%%%%%%%%%%%%%%%%%%%%%%%%%

The step towards a theoretical modeling of ultrafast processes at atomistic scale is limited by the \textit{accuracy} problem intrinsic to all approximate solutions to the time-dependent many-body molecular Schr\"odinger equation. Approximations in \textit{ab initio} molecular dynamics occur in different flavors.% simulations are twofolds.  
First, one can transform the original system Hamiltonian into a new form more suited for a numerical solution.
This is for instance the case in DFT where the complex interacting Hamiltonian
is mapped (in an in principle exact way) into a system of noninteracting particles subject to the action of a compensating local external potential (the exchange-correlation (xc) potential) for which we need approximations.
More extreme approximations of the Hamiltonian are obtained in the so called semi-empirical methods. 
Other approaches use the exact Hamiltonian combined with approximated expansions of the many-electron wavefunctions. These methods comprise configuration interaction (CI) approaches, multi-configurational self-consistent field (CASSCF) with or without corrections to include dynamical correlation (M{\o}ller--Plesset perturbation), coupled cluster (CC) and many other post-Hartree Fock methods. The situation becomes even more complicated when in addition to the approximations to the electronic part of the Schr\"odinger equation, one also considers nuclear quantum effects, which are of particular relevance in nonadiabatic dynamics~\cite{curchod2013trajectory}.

The second important source of approximations is linked to numerics. Despite the enormous 
increase of computational resources in the last decades, the unfavorable scaling associated to many electronic structure approaches hampers an adequate sampling of the electronic wavefunction space both in the ground and excited states. When coming to \textit{ab initio} dynamics, the situation becomes even more critical since the numerical costs associated to the evaluations of energies and forces are multiplied by the number of time steps required to describe the process of interest. In general, ground state simulations based on DFT are limited to tens of picoseconds while nonadiabatic simulations can rarely be pushed over one ps. These approximations rise inevitably the issue about the \textit{predictivity} of \textit{ab initio} molecular dynamics simulations. Unfortunately, there is no way to predict \textit{a priori} the effects of the approximations introduced at the level of the Hamiltonian or for the description of the many-electron wavefunctions and therefore a number of tests has to be performed to assess the quality of the the xc-functional in DFT, the many-electron wavefunction representation in post-Hartree Fock approaches, and the basis set used for the expansion of the molecular orbitals. These quality checks are usually performed using very accurate single point calculations as references (when affordable) or, even better, experimental results.

A word of caution is however required when comparing experimental and numerical results. Due to the difficulties in controlling the experimental conditions the match with the numerical calculations can often be problematic: effects related to the environment, sample size, pressure and temperature are difficult to reproduce in a calculation and will always give a margin of uncertainty.

%%%%%%%%%%%%%%%%%%%%%%%%%%%%%%%%%%%%%%%%%%%%%%%%%%%%%%%%%%%%%%%%%%%%%%%%%%%%%%%%
  \subsection{Constrained DFT and redox processes}\label{sec:constrdft}
%%%%%%%%%%%%%%%%%%%%%%%%%%%%%%%%%%%%%%%%%%%%%%%%%%%%%%%%%%%%%%%%%%%%%%%%%%%%%%%%
Ultrafast charge transfer processes such as electron transfer (ET) processes can often be efficiently studied in a `static' picture using the framework of Marcus theory \cite{Kaduk_2012a}, which describes ET kinetics in terms of diabatic parabolic curves that can be determined using ground state \textit{ab initio} MD simulations. Extensions to the non-Marcus regime have been also extensively studied~\cite{blumberger2004electronic, vuilleumier2012} and they can be easily implemented as corrections to the linear response (Marcus) picture.

The simulation of Marcus diabatic curves implies the separation of the full ET problem in its two main constituents, namely the oxidation and the reduction half-reactions. The main advantage of this approach compared to the simulation of the entire process in a unique computational setups relies in the fact that DFT with semilocal functionals has notably the tendency of over-delocalizing the electronic distribution in charge separated states. This problem can be partially solved using more advanced functionals following Perdew's ranking~\cite{ladder} (Jacob's ladder of functional). In particular, range-separated hybrids as CAM-B3LYP~\cite{cam} and $\gamma$-tuned functionals~\cite{LR_correction_scheme,Stein} often provide a good description of the charge separated states in the ground and excited states.
%To overcome this problem, one solution consists in the introduction of charge constraints (as discussed in section~\ref{}), while the other relies in the modeling of the constituent half-reactions. 
Within this approach, we instead only consider redox half-reaction of the type
\begin{equation}
  R \rightarrow O +e^-
  \label{Eq:RtoO}
\end{equation} 
where the reduced specie (R) is put in contact with an electron reservoir at constant chemical potential~\cite{sprik02}, which absorbs the emitted electron in the formation of the oxidized state. In this process, the number of electrons in the system varies from $N$ in the reduced state to $N-1$ in the oxidized state. 

In practice, the method is based on the observation that the thermal distribution of vertical energy gap of the $R \rightarrow O +e^-$ and $O + e^- \rightarrow R$ reactions obtained through MD sampling contain all relevant information needed to compute the reaction free energy for the underlying electron transfer process~\cite{sprik02,blumberger2004electronic}. For the half reaction in (\ref{Eq:RtoO}) this amounts to the free energy of oxidation $\Delta A$. The relation between the vertical energy gap and the oxidation free energy $\Delta A$ is particularly simple when the response of the solvent to ET is linear (meaning in the Marcus regime). The redox free energy is then directly obtained as the mean vertical energy gap~\cite{sprik02}
\begin{equation}
  \Delta A= \frac{1}{2}(\langle \Delta E_0\rangle_R + \langle \Delta E_0\rangle_O  )
\end{equation} 
where the brackets $\langle \dots \rangle_{O/R}$ represent the ensemble average of the reduction, respectively oxidation, energy gaps sampled on the ground state PES. Similarly, the difference of mean energy gaps can be shown to give an estimate of the reorganization free energy
\begin{equation}
  \lambda= \frac{1}{2}(\langle \Delta E_0\rangle_R - \langle \Delta E_0\rangle_O ) \, .
\end{equation} 
These gap relations for the redox free energy and the reorganization free energy are remarkably powerful, enabling one to estimate these quantities from two equilibrium molecular dynamics runs. Similar expressions also hold for the reaction and the reorganization free energies that appear in the Marcus gap law for the free energy of activation.

So far about the thermodynamics of the ET process. In order to access kinetic properties one needs to compute the activation barriers and the nonadiabatic coupling elements between the ground and the first excited states potential energy surfaces. 
To this end, one can perform constrained MD simulations to enrich the sampling of the regions of strong coupling between the two diabatic states using the grand canonical potential energy obtained from a linear superposition of the oxidized and reduced states~\cite{sprik02,blumberger05} or a charge constrained approach~\cite{blumberger05}. 
At the avoidded crossings, the nonadiabatic couplings are then computed using excited states approaches such as linear response TDDFT (LR-TDDFT). 
From the free energy barrier, $\Delta A^{\dagger}$, and the free energy splitting, $\Delta A_{RO}$, an estimate of the ET kinetic rate can be derived.

%%%%%%%%%%%%%%%%%%%%%%%%%%%%%%%%%%%%%%%%%%%%%%%%%%%%%%%%%%%%%%%%%%%%%%%%%%%%%%%%
  \subsection{Methods based on Density Functional Theory}\label{sec:dft}
%%%%%%%%%%%%%%%%%%%%%%%%%%%%%%%%%%%%%%%%%%%%%%%%%%%%%%%%%%%%%%%%%%%%%%%%%%%%%%%%
The combination of powerful electronic structure methods such as DFT and TDDFT with an adequate description of the nuclear dynamics has allowed the development of a family of adiabatic and nonadiabatic molecular dynamics schemes for the investigation of the quantum dynamics of realistic systems (at atomistic scale) embedded in their physical environment.
As outlined in section~\ref{sec:QD}, the main challenge in the derivation of these quantum dynamics approaches consists in the design of a suited representation of the nuclear quantum dynamics. In fact, while the quantum mechanical description of the electron dynamics can be correctly reproduced with the help of different electronic structure approaches (in particular DFT and TDDFT), the description of the combined electron-nuclear dynamics is still posing enormous theoretical and computational challenges.
We have therefore to relay on different approximations that range from the classical treatment of the nuclei as point charges, to semiclassical and mixed quantum-classical representations, to more accurate (but also computationally still intractable) solutions based on nuclear wavepacket dynamics or Bohmian quantum trajectories (within the quantum hydrodynamics formalism). 
%A brief discussions on these last methods will be given in the following sections together with the description of possible future developments (see also my most recent publications in Refs.~\cite{tavernelli2013ab} and~\cite{tavernelli2015nonadiabatic}).

More specifically, in the last years we observed an extensive development of DFT and TDDFT based electronic structure and  molecular quantum dynamics techniques. 
The main reasons for this phenomenon are: 
\begin{enumerate}
  \item the ``densityfunctionalization" of time-dependent Kohn-Sham propagation scheme combined with nuclear dynamics (Ehrenfest dynamics) \cite{Li_2005a,Andrade_2009a}; 
  \item the implementation of nonadiabatic mixed quantum-classical molecular dynamics schemes based on Tully's trajectory surface hopping (TSH)~\cite{tapavicza07, tavernelli09, tavernelli09b}.  To this end, surface hopping probabilities were accurately computed using solely density-based quantities, and excited state energies and forces were derived from LR-TDDFT;
  \item the ``densityfunctionalization" of nonadiabatic vectors (important for the detection of conical intersections on the potential energy surfaces and for the rescaling of the nuclear velocities after each surface hop) and their implementation~\cite{tapavicza07, tavernelli09, tavernelli09b,tavernelli10b, curchod2013trajectory} in the CPMD software package~\cite{cpmd}; 
  \item the development of a coupling scheme for the inclusion of external time-dependent electric fields (in particular laser fields)~\cite{tavernelli2010mixed}.
  \item the derivation of spin-orbit couplings (SOC) within TDDFT and their combination with TSH dynamics for the investigation of inter-system crossing events~\cite{decarvalho2014derivation,carvalho17}.
\end{enumerate}

%%%%%%%%%%%%%%%%%%%%%%%%%%%%%%%%%%%%%%%%%%%%%%%%%%%%%%%%%%%%%%%%%%%%%%%%%%%%%%%%
  \subsection{Applications of ab-initio methods}\label{sec:abinitappl}
%%%%%%%%%%%%%%%%%%%%%%%%%%%%%%%%%%%%%%%%%%%%%%%%%%%%%%%%%%%%%%%%%%%%%%%%%%%%%%%%

%%%%%%%%%%%%%%%%%%%%%%%%%%%%%%%%%%%%%%%%%%%%%%%%%%%%%%%%%%%%%%%%%%%%%%%%%%%%%%%%
    \subsubsection{Ultrafast dynamics of photoexcited metal complexes at atomistic resolution}
\begin{figure}[h]
  \centering
  \includegraphics[width=0.48\textwidth]{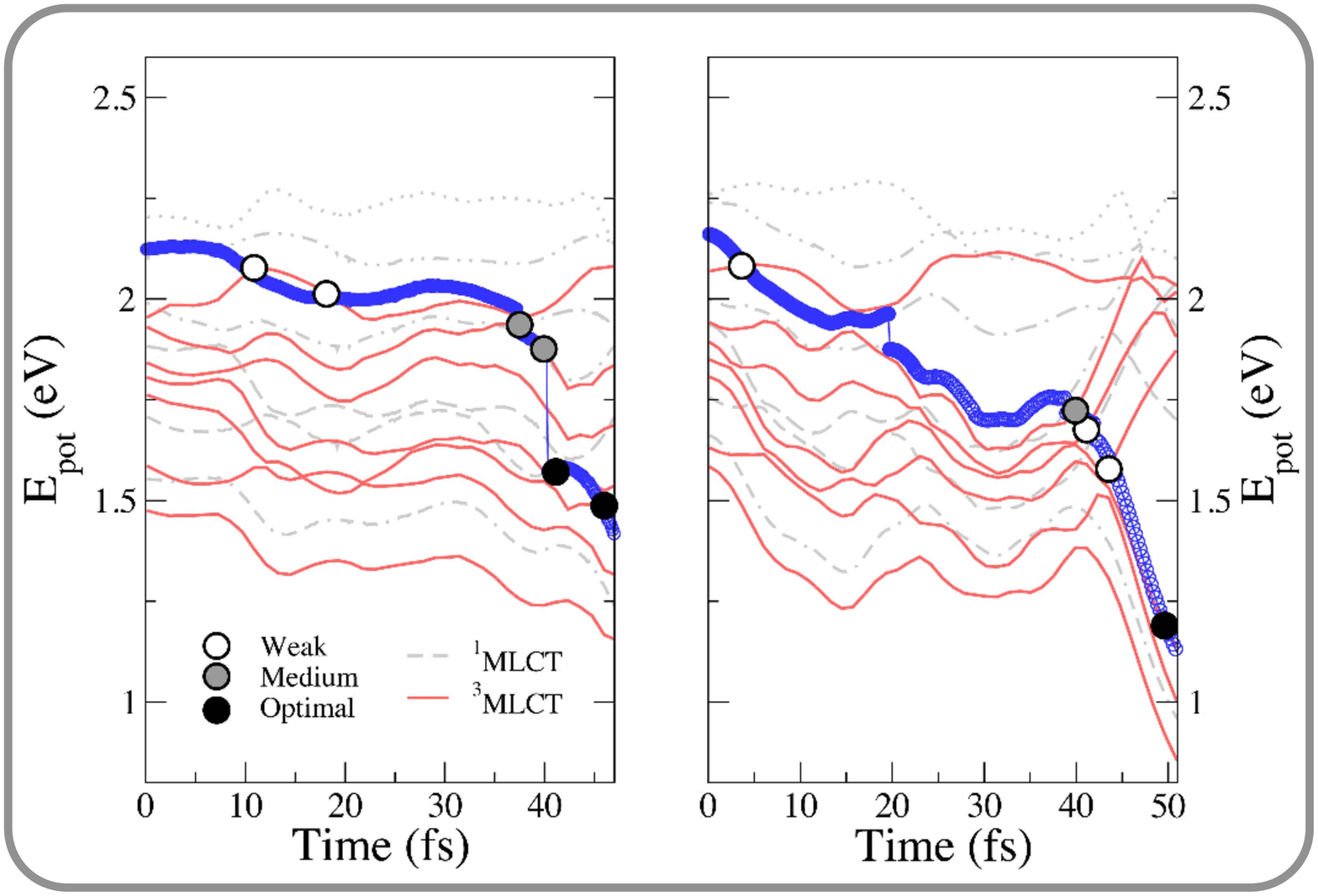}
  \caption{Nonadiabatic molecular dynamics of [Ru(bpy)$_3$]$^{2+}$ in water. The two panels show the time series of the most relevant excited state energies for two representative trajectories. Gray: singlet states (7 in total); red: triplet states (7 in total). The force state (the one that drives the dynamics) is highlighted with blue circles. The crossing points between singlet and triplet states are shown with filled circles. The corresponding magnitude is color coded: white=weak, gray=medium, and black=large spin orbit coupling intensity. Reprinted from \cite{tavernelli2011nonadiabatic}, Copyright 2011, with permission from Elsevier.}
  \label{MLCT:relax1}
\end{figure}

%%%%%%%%%%%%%%%%%%%%%%%%%%%%%%%%%%%%%%%%%%%%%%%%%%%%%%%%%%%%%%%%%%%%%%%%%%%%%%%%
Tris(bi-pyridine) metal compounds are the prototypes of the metal-to-ligand charge transfer (MLCT) complexes. 
The photophysics of these systems consists of a stimulated singlet-to-singlet excitation of one electron from the central metal atom into the ligand system ((bpy)$_3$), followed by an ultrafast intersystem crossing to the triplet states. 
% Solvent.
While in the gas phase the excited electron distributes uniformly among the three ligands, in solution the situation can be rather different depending of the nature of the solvent.
In fact, the anisotropic arrangement of the first solvation shells together with the ultrafast reorientation of the individual dipoles of the solvent molecules in the bulk cause a clear localization of the excited electron on mainly one (but sometimes two) ligands. 
In the case of water~\cite{Moret2009a, Moret2010}, the first solvation shell is organized into a linear chain of hydrogen bonded water molecules arranged along the grooves between the ligands. 
Through the rotation around the O-H bonds involved in this chain, the individual water dipoles can rearrange in the fs time-scale providing an ultrafast mechanism for charge stabilization upon photoexcitation. 
%Clearly, the level of details needed for the description of this organized solvation structure requires the use of an atomistic representation of the solvent. 
%A coarse-grained or implicit description of the solvent will miss some of the important effects induced by the solvent dynamics and therefore are not appropriate for an accurate account of the detailed structural reorganization of the solvent that influence the dynamics of the MLCT complex in the sub picosecond time-scale.
%the dynamics
Using TDDFT-based TSH dynamics, we were able to investigate the ultrafast intersystem crossing (ISC) dynamics which follows the photoexcitation of Rutheniun(II)this(bipyridine) in water~\cite{tavernelli2011nonadiabatic}. 
After photon absorption the system is directly excited into the brightest singlet MLCT state, from which it relaxes through %internal conversion into the lower lying singlet states . 
internal conversion into the lower lying manifold of singlet states (Figure~\ref{MLCT:relax1}).
%%%%%%%%%%%%%%%%%%%%%%%%%%%%%%%%%%%%%%%%%%%%%%%%%%%%%%%%
% Figure 7
%%%%%%%%%%%%%%%%%%%%%%%%%%%%%%%%%%%%%%%%%%%%%%%%%%%%%%%%

In agreement with experiments performed on similar compounds~\cite{Chergui_science2009}, the first part of the dynamics ($\sim 100$ fs) is characterized by a ultrafast energy relaxation, which give rise to a series of ISCs between singlet and triplet states. The dynamics of this process can be easily rationalized by means of our nonadiabatic calculations in explicit solvent (TDDFT/MM TSH dynamics): after excitation, the electron transferred to the ligand system (which is mainly localized on a single bi-pyridine) is stabilized by the dipole reorientation of the water molecules in the first solvation shell. 
This process is made possible by the one-dimensional chain of water molecules intercalated between the ligands, which are connected to each other by a single hydrogen bond.  Due to the size of the spin-orbit coupling between lower lying singlet and triplet MLCT states, the $^1$MLCT $\rightarrow$ $^3$MLCT transition occurs in all TSH trajectories in about 50 to 80 fs.
Fast ISCs have important implications in nanotechnologies, as for instance in the realization of efficient dye-sensitized solar cell devices. In fact, large ISC rates allow for an efficient population transfer into triplet states, protecting the dyes from electron-hole back recombination and therefore favoring charge separation at the interface between the donor (dye) and the acceptor (the semiconductor nanoparticles).
%In this case, the use of accurate \textit{ab initio} molecular dynamics calculations gives important insights for the interpretation of the experimental results and allows for a detailed explanation of the intra- and inter-molecular mechanisms governing this complex photophysical process.

The subsequent long-term dynamics of the triplet MLCT complex is followed using adiabatic \textit{ab initio} MD simulations~\cite{Moret2010}. In this case, simple adiabatic Car-Parrinello or Born-Oppenheimer dynamics in the lowest triplet state give an accurate description of the spin density dynamics in the ligand system (figure~\ref{MLCT:trpMD}).
In particular, we observe the localization of the photo-electron mainly on a single ligand with attempts to partially delocalize to a second one. This fluxual asymmetric MLCT configuration (which differs from the homogeneous distribution observed in gas phase) is stabilized (by about 1 V~\cite{Moret2010}) through the orientation of the water dipoles of the bulk solution.

%%%%%%%%%%%%%%%%%%%%%%%%%%%%%%%%%%%%%%%%%%%%%%%%%%%%%%%%%%%%%%%%%%%%%%%%%%%%%%%%
    \subsubsection{Electron transfer in photoexcited azurin}
%%%%%%%%%%%%%%%%%%%%%%%%%%%%%%%%%%%%%%%%%%%%%%%%%%%%%%%%%%%%%%%%%%%%%%%%%%%%%%%%
%1JZF
%Pseudomonas aeruginosa Oxidized Azurin(Cu2+) Ru(tpy)(phen)(His83)
%1JZG
%Pseudomonas aeruginosa Reduced Azurin (Cu1+) Ru(tpy)(phen)(His83) (2001) J Am Chem Soc 123 11623-1163
In proteins, long-range electron transfer occurs through a series of electron-tunneling pathways along secondary structure elements ($\beta$-strands), aromatic side chains and protein-solvent interphases. Interestingly, experimental~\cite{Regan1995, Shih1760} and theoretical studies~\cite{curry_pathways_1995,Gray2003,Beratan2009,warren_azurin_2013} point towards the existence of multiple concurring ET paths through the protein matrix, which are fragmented into through-bond and through-space tunneling components. In particular, there is a wide consensus that aromatic residue side chains can mediate the ET process in the hydrophobic protein core. However, the protein matrix does not form a continuum and therefore tunneling through space and in particular along hydrogen-bond networks is probably one of the most important rate limiting steps of the overall transport process.

\begin{figure}[h]
  \centering
  \includegraphics[width=0.48\textwidth]{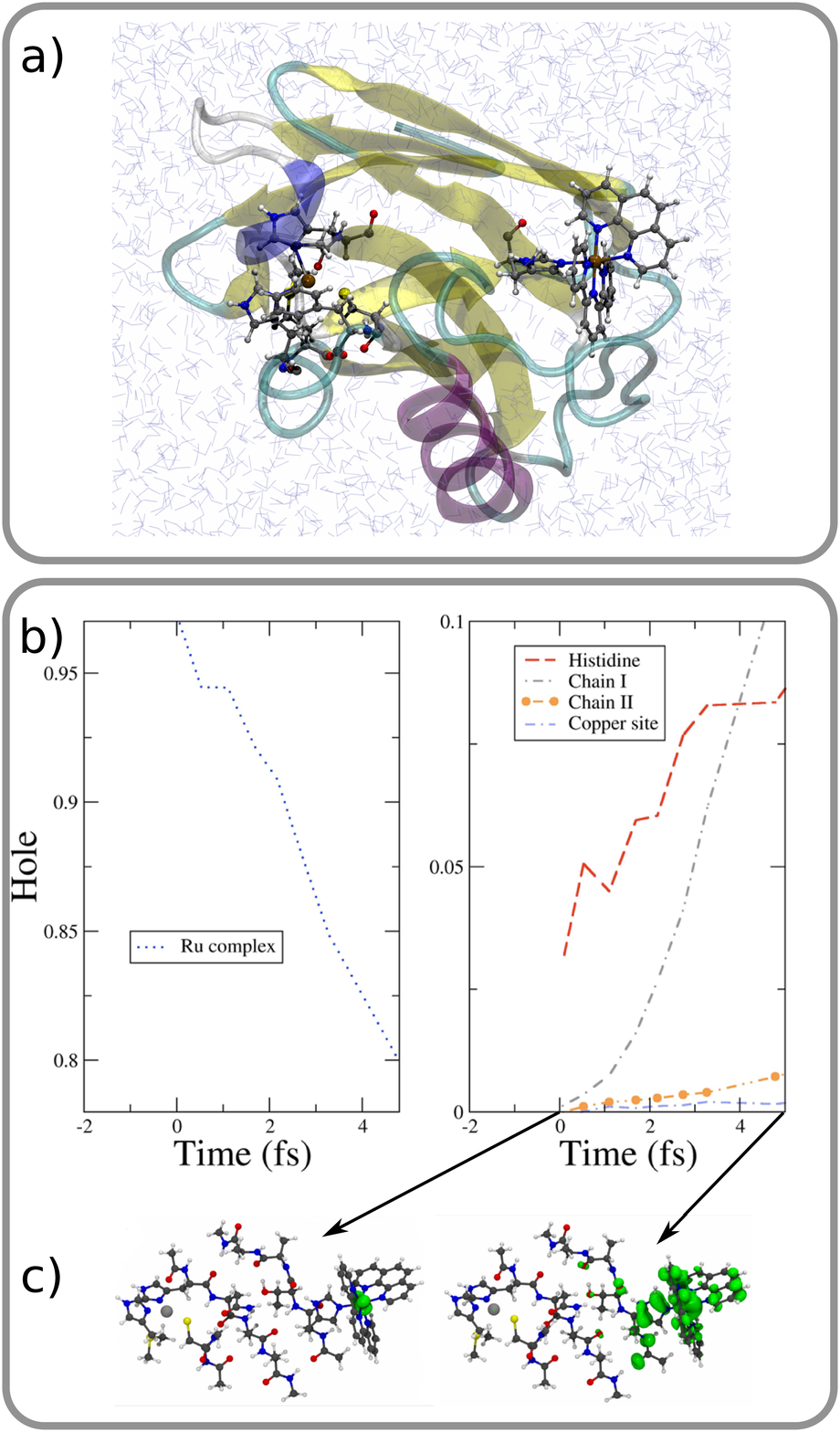}
  \caption{a) Representation of the QM/MM setup used for the simulation of the HT process in azurin. The protein is shown in the `cartoon' representation, which highlights the secondary structure elements, while the acceptor and donor centers are given in the ball-and-stick representation. The water is described with the simple point charge model and is represented in violet. b) Time series with the evolution of the hole density partitioned in five main groups (i.e., integrals of $\rho^{\textrm{hole}}(r)$ over different molecular fragments). Left panel: the Ru-complex; Right panel: the His(83), the first $\beta$-sheet strand (Chain I close to the Ru ion), the second $\beta$-sheet strand (Chain II close to the Cu ion), and the copper binding site. (c) Left: initial position of the electron hole centered at the Ru ion; Right: distribution of the electron hole on the quantum sub-system after 4.5 fs of dynamics. The simulations were performed using the TDDFT-based Ehrenfest dynamics module as implemented in the CPMD software package~\cite{cpmd}. }
  \label{fig:azurin}
\end{figure}

\begin{figure}
  \centering
  \includegraphics[width=0.48\textwidth]{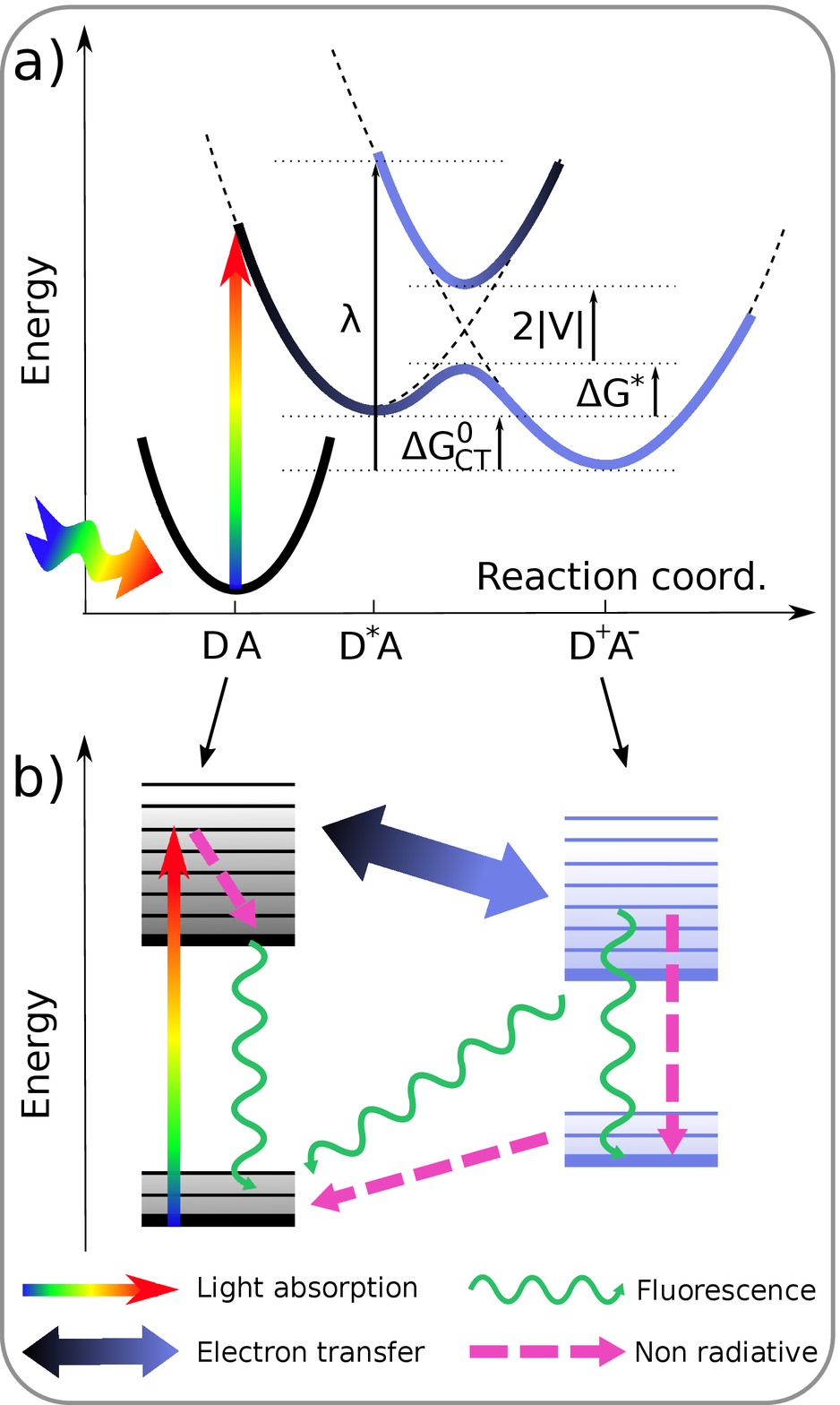}
  \caption{(a) Schematic picture of the photoinduced electron transfer energetics. $\Delta G^0_{CT}$ is the Gibbs free energy change for the electron transfer reaction, $\lambda$ the reorganization energy. Dashed black lines and thick color lines respectively indicate diabatic and adiabatic potential energy surfaces of reactants and products. (b) Structure of electronic (thick horizontal lines) and vibrational states (thin horizontal lines). The channels for charge transfer, radiative and non-radiative recombination and relaxation are also shown.}
  \label{fig:marcus}
\end{figure}

\begin{figure}[ht]
  \centering
  \includegraphics[width=0.48\textwidth]{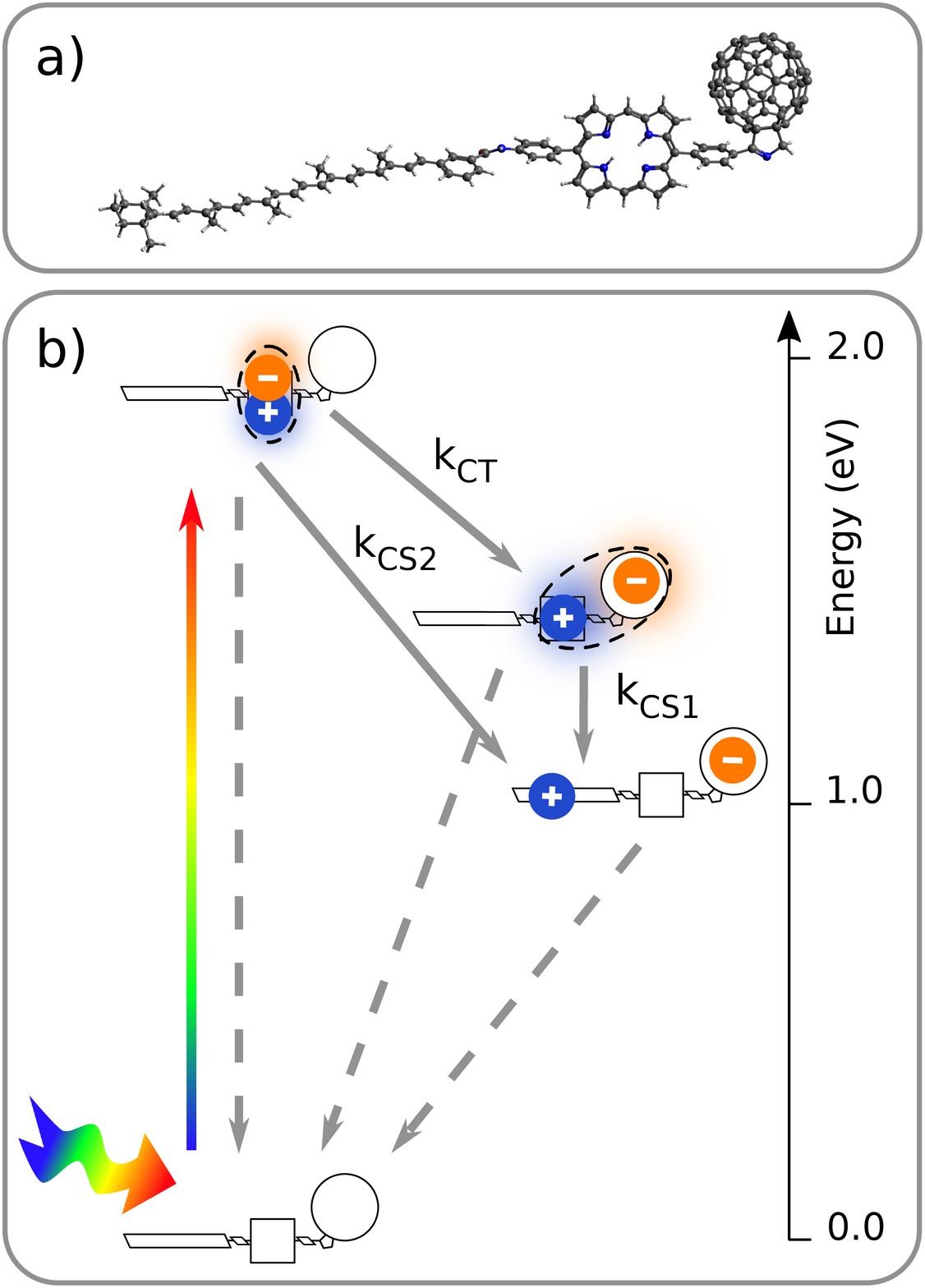}
  \caption{(a) linear structure of a carotenoporphyrin-fullerene triad \cite{Baruah_2006a,Spallanzani_2009a}. (b) simplified excited states and interconversion pathways in the CPC$_{60}$ triad. Triplet states are omitted for simplicity. Possible values of the relaxation rates in 2-methyltetrahydrofuran solution are \cite{Liddell_1997a}: $k_{CS1} \approx 10^{10}$ s$^{-1}$ and $k_{CS2} \approx 10^{11}$ s$^{-1}$.}
  \label{fig:triad_gust}
\end{figure}

\begin{figure}[hb]
  \centering
  \includegraphics[width=0.48\textwidth]{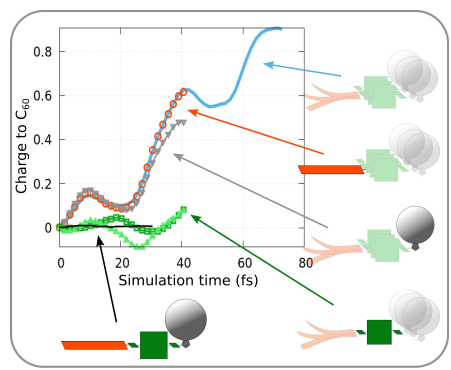}
  \caption{TDDFT simulation of sub-100 fs charge separation in the CP$_{60}$ triad under different nuclear constraints \cite{Rozzi_2013a}. Clamping nuclei in groups leads to totally different dynamics. Blocking the motion of the atoms of the linker between porphyrin and fullerene stops the charge accumulation at the acceptor. The light blue thick line refers to the free molecule; orange circles to clamped carotene; gray triangles to clamped C$_{60}$; dark green squared to clamped porphyrin; light green triangles to clamped linker ring between the porphyrin and the C$_{60}$; black thin line to all clamped nuclei.}
  \label{fig:triad_carlo}
\end{figure}

\begin{figure}[h]
  \centering
  \includegraphics[width=0.48\textwidth]{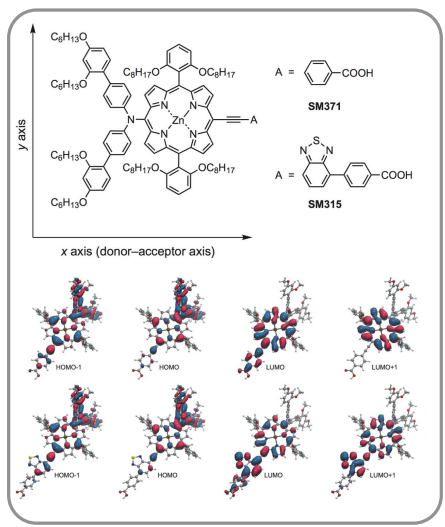}
  \caption{Upper panel: structures of two porphyrin-based dyes (coded SM371 and SM315). They both feature a porphyrin core and a bulky bis(2',4'-bis(hexyloxy)-[1,1'-biphenyl]-4-yl)amine donor group; the acceptors are different and are responsible for their photo-physical properties. Lower panel: frontier orbitals characterizing in the low energy excitations in SM371 (first line) and SM315 (second line). The main transitions contain a large HOMO-LUMO component, which in the case of SM315 is associated with a strong charge separation character. Reprinted by permission from Macmillan Publishers Ltd: Nature Chemistry \cite{mathew2014dye}, Copyright 2014.}
  \label{Fig:Porphyrin}
\end{figure}

\begin{figure}
  \centering
  \includegraphics[width=0.48\textwidth]{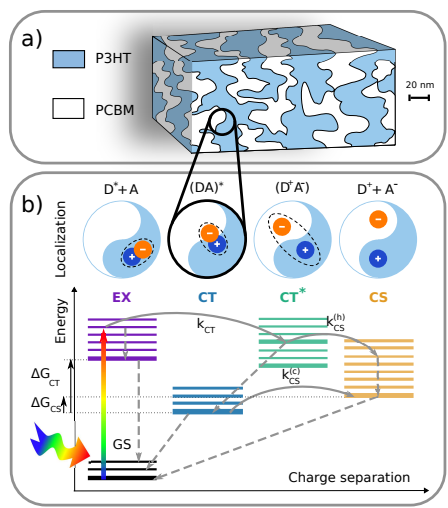}
  \caption{a) Schematic view of a typical heterojunction. b) Schematic view of excited states involved in the charge separation at a donor--acceptor bulk heterojunction interface. GS indicates the ground state; EX the exciton states localized on the donor; CT the interfacial charge transfer state; CT$^*$ excited charge-transfer states; CS the charge-separated state. Curved continuous lines indicate charge separation paths; broken lines relaxation or recombination paths. $k_{CT}$ is the rate constant for the EX to CT crossover; $k^{(c)}_{CS}$ and $k^{(h)}_{CS}$ respectively indicate the charge separation rate constants for cold and hot excitons. Thick lines indicate electronic energy levels; thin lines vibrational levels. Triplet states are omitted for simplicity.}
  \label{fig:views}
\end{figure}

\begin{figure}
  \begin{center}
  \includegraphics[width=0.45\textwidth]{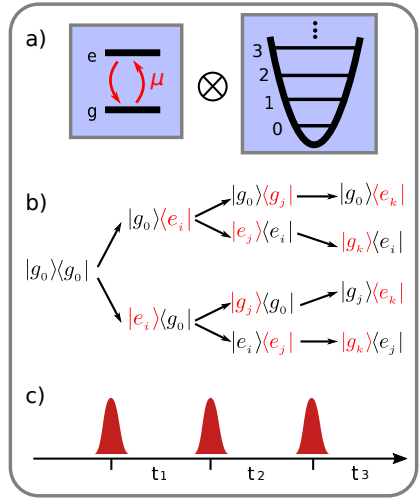}
  \end{center}
  \caption{a) Prototypical model, formed by a two-level system and an harmonic oscillator (vibrational mode), coupled to each other by the Hamiltonian in (\ref{eqFT10}). b) Schematic view of the possible pathways that contribute to the third-order polarization, starting for a system initialized in the ground state $|g_0\rangle$. Only final terms of the kind $| g_i \rangle\langle e_j |$ have been included, the others can be obtained by replacing each operator in the pathway with its Hermitian conjugate. Each laser pulse changes either the ket or the bra. Kets and bras in red are the ones affected by the field. The subscripts denote the phonon number in the displaced oscillator basis, and can take in principle arbitrary values. c) Sequence of two pump pulses and one probe pulse used in the 2D spectroscopy, with the corresponding waiting times.}
  \label{figFT01}
\end{figure}

\begin{figure}[h]
  \centering
  \includegraphics[width=0.48\textwidth]{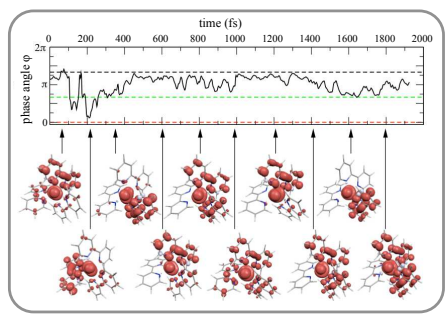}
  \caption{Time evolution triplet state spin density during the dynamics at 300 K in water (QM/MM setup). The dashed lines at $\phi$ angles of 0, $2\pi/3$, $4\pi/3$ represent the direction of vectors pointing from the center of ligands 1, 2, and 3, respectively, to the ruthenium atom. They correspond to the localization of the unpaired electron on ligand 2 (red), 3 (green), 1 (black), respectively. Bottom: spin densities computed from snapshots of the QM/MM trajectory. Figure adapted with permission from \cite{Moret2010} John Wiley \& Sons. Copyright \textcopyright\ 2010 WILEY-VCH}
  \label{MLCT:trpMD}
\end{figure}

The hole-transfer (HT) dynamics in the mutated Pseudomonas Aeruginosa azurin protein has been investigated using the TDDFT-based Ehrenfest dynamics approach described in section~\ref{subsub_MQC}. 
At the surface of the protein a Ru(tpy)(phen)(His83) complex has been linked to the His83 residue (tpy = trisbipyridine, phen = phenentroline, His = Histidine). The electron/hole transfer from the Copper to Ru metal center occurs over a distance of $\sim 25$~\AA~  through a double-strand `protein bridge'. The system (AzRu) is treated at a (TD)DFT/MM level of theory (the preparation of the system follows closely the one reported for a similar system in Ref.~\cite{cascella07}). The Ru complex at the surface, the linking residue (His83), the copper binding site and the protein backbone bridging the two metallic centers are included in the DFT part, while the rest of the system (protein and the solvent) is modeled using a classical force field (Amber FF~\cite{amber9}). The dynamics is initiated with the oxidation of the Ru-center (Cu(I)/Ru(III)). This is achieved by removing an electron from the highest occupied electron at the ruthenium ion. After the ionization, the hole created at the Ru-complex is transferred to the copper ion leading to the final oxidation state Cu(II)/Ru(II).
As a measure of the hole transfer process as a function of time, we monitor the evolution of the system spin density, 
$\rho^{\textrm{hole}}(r)=\rho^{\alpha}(r)-\rho^{\beta}(r)$ and in particular the values of the integrals of $\rho^{\textrm{hole}}(r)$ evaluated for different fragments (see figure~\ref{fig:azurin}, panel b)). The quantum subsystem is composed of 191 atoms placed between the two redox centers. The propagation of the Kohn-Sham orbitals was performed with a time step of $10^{-3}$ fs using the PBE exchange and correlation functional~\cite{perdew96} within the adiabatic approximation for the TDDFT kernel. All simulations were performed in the canonical ensemble at 300 K.

%%%%%%%%%%%%%%%%%%%%%%%%%%%%%%%%%%%%%%%%%%%%%%%%%%%%%%%%
% Figure 9
%%%%%%%%%%%%%%%%%%%%%%%%%%%%%%%%%%%%%%%%%%%%%%%%%%%%%%%%

The fulll HT process occurs in the ns time scale and therefore it is too slow to be monitored using the Ehrenfest dynamics approach. In order to speed up the hole-transfer (HT) dynamics, the system was therefore initially equilibrated (with classical MD) with a set of charges that correspond to the Cu(II)/Ru(II) oxidation state. This creates a driving force for the hole to move from ruthenium to the copper ion. In fact, a wavefunction optimization at fix nuclear positions leads directly to the product state (Cu(II)/Ru(II)). Despite this bias, the kinetics for the HT process is still too slow and therefore we can only access the early stages of the dynamics (first 5 fs). As shown in figure~\ref{fig:azurin} (panel b), this time window is enough to capture the first few percents of the HT process, giving indications about the nature of the mechanism. 

In particular,  an ultrafast distribution of the hole in the ligand system of the Ru-complex is first observed (figure~\ref{fig:azurin}, panels (b) and (c), followed by a diffusion into the (covalently bounded) first $\beta$-sheet strand. At this point, the HT can only proceed via the tunneling through hydrogen bonds between antiparallel $\beta$-strands leading to a slow down of the process. In the time scale of our simulation ($\sim 5$ fs) only a minor fraction ($< 1$\%) of the hole reaches the copper atom. Longer simulation times and more detailed analysis will be needed to shed full light on this  fundamental process.

\section{Outlook}

% Ultrafast
Charge transfer reactions have been studied for long time by both chemists and physicists, especially addressing the reaction energetics and the kinematics.  The recent advent of lasers able to generate femtosecond (or attosecond) pulses, and the corresponding development of high time-resolution spectroscopies have opened an entire new window of opportunities for investigation. The motion of microscopic charge carriers can be monitored in real time in a wide variety of systems, from bacterial photosynthetic complexes to molecular or supramolecular aggragates, to classical, or nano-structured solid state devices. This opportunity has in turn stimulated the need for corresponding theoretical advances, able to describe and rationalize the results of time-resolved observations, connect the observed quantities to the microscopic quantum dynamics of the systems, predict the properties of new materials, and discover new physical regimes, phases and phenomena.

% Molecular ET
In section \ref{sec:proto} we have reviewed the features of charge separation in few prototypical cases, representative of the aforementioned categories. Molecular charge separation is usually well understood in its general features, and rationalized within a simple, yet very powerful framework --Marcus' theory-- capable of providing rate constants for different ranges of reorganization energy differences. This holds, provided that parabolical potential energy surfaces of nuclear motion are known. The situation in the ultrafast regime is more complex, and the coupling between electrons and nuclei is profitably probed by means of ultrashort laser pulses, which are able to unravel coherent collective behaviours not included in Marcus' picture. Although a fully quantum description has not been reached, except from very small molecules and model systems, the theoretical description based on first principles methods has reached a good level of maturity and reliability in  predicting charge separation properties, by means of the quantum-classical, wave-packet and trajectory based methods. These are reviewed in section \ref{sec:QD}, and a selection of succesful applications is reported in section \ref{sec:supra} and \ref{sec:abinitappl}.

% Nanoscale ET
The situation is way more problematic when we turn to nanostructured systems, such as those illustrated in Secs. \ref{sec:dye} and \ref{sec:bulk}. Here the role of the mesoscopic structure of the samples renders them less reproducible and introduces new layers of low energy interactions, which might alter the microscopic local physics, make long-range interactions important, such that the role of delocalized states must be taken into account. In these situations also experimental probes relying on macroscopically averaged quantities may provide contrasting results (see the discussion in section \ref{sec:hot}). This is probably the reason why the mechanisms powering new generation photovoltaic materials, which mostly rest on nano-structuring, remain difficult to understand completely. In particular, in section \ref{sec:cur} we have sketched the recent debate about the nature of states mediating efficient charge separation in bulk heterojunctions. The picture of tightly bound molecular excitons collides with the time scale and efficiency of the process. Some authors suggested that exciton excess energy and/or delocalization could bypass trap states skipping relaxation to the lowest charge transfer state. Other authors and data are instead in support of a multi-step process. None of the proposed solutions to this problem have reached a general consensus yet, and also the link between the ultrafast regime and common macroscopic indicators of device efficiency remain uncertain.

% From delocalization to coherence
The issue of charge delocalization is bound to a similar problem that appeared in a different class of charge and energy-transferring systems, i.e. natural photosynthetic complexes, and leads to a different approach. Due to the large size of the these systems, atomistic simulations of their overall state are still off the table. Multiscale and QM/MM models have been extensively used in this field (for a recent review see Ref. \cite{Curutchet_2017a}), but substantial information about the role of coherence may be obtained by  pairing 2D spectroscopy and model Hamiltonians. In this way, the existence of different types of coherence (electronic, nuclear, vibronic) was hilighted and discussed. These topics are reviewed in sections \ref{sec:coh} and \ref{sec:nuc}. Some findings turned out to be relevant also for non-biological systems, and further raised questions about the importance of truly quantum features in charge separation and energy transfer processes. 

The concepts of coherence and entanglement have been quite thoroughly explored by physicists in different contexts, ranging from optical and atomic to solid-state systems. However, to our surprise, there has been relatively little cross-feeding so far. This is mainly why we have decided to include in Secs. \ref{sec:quantcoh} and \ref{sec:thspect} a review of methods and indicators of coherence and entanglement, although their applications are so far rather sparse and limited to model systems that include only few degrees of freedom. In fact, we believe that these quantities will turn out to be useful in future developments of photovoltaics, along with a higher degree of hybridization between theoretical tools developed in the fields of quantum information processing and ab initio simulations.

% Challenges
In conclusion, matching theory and experiments in the field of photoinduced charge separation means complementing information. The overall motion of charges, their time-dependent localization, and the energetics of excited states can be often accurately extracted from different experimental techniques (see section \ref{sec:exp}). However, revealing the nature and the interactions of the corresponding excited states requires  reliable theoretical and numerical models. While the electronic structure and the coupled nuclear dynamics of molecular and bulk systems are quite thoroughly known, and are now accessible also by first principles methods, in the case of nanostructured materials much work still has to be done. Furthermore, a fully quantum method that can be applied to ultrafast charge separation phenomena in nanostructured systems is still to come.

%%%%%%%%%%%%%%%%%%%%%%%%%%%%%%%%%%%%%%%%%%%%%%%%%%%%%%%%%%%%%%%%%%%%%%%%%%%%%%%%
\section*{Acknowledgements}
%%%%%%%%%%%%%%%%%%%%%%%%%%%%%%%%%%%%%%%%%%%%%%%%%%%%%%%%%%%%%%%%%%%%%%%%%%%%%%%%
CCAR acknowledges financial support from FP7-NMP-2011-SMALL-5 ``CRONOS" (grant No. 280879-2) and FP7-MC-IIF ``MODENADYNA" (grant No. 623413).

%%%%%%%%%%%%%%%%%%%%%%%%%%%%%%%%%%%%%%%%%%%%%%%%%%%%%%%%%%%%%%%%%%%%%%%%%%%%%%%%
\section*{Figures}
%%%%%%%%%%%%%%%%%%%%%%%%%%%%%%%%%%%%%%%%%%%%%%%%%%%%%%%%%%%%%%%%%%%%%%%%%%%%%%%%

%%%%%%%%%%%%%%%%%%%%%%%%%%%%%%%%%%%%%%%%%%%%%%%%%%%%%%%%
% Figure 1
%%%%%%%%%%%%%%%%%%%%%%%%%%%%%%%%%%%%%%%%%%%%%%%%%%%%%%%%

%%%%%%%%%%%%%%%%%%%%%%%%%%%%%%%%%%%%%%%%%%%%%%%%%%%%%%%%
% Figure 2
%%%%%%%%%%%%%%%%%%%%%%%%%%%%%%%%%%%%%%%%%%%%%%%%%%%%%%%%

%%%%%%%%%%%%%%%%%%%%%%%%%%%%%%%%%%%%%%%%%%%%%%%%%%%%%%%%
% Figure 3
%%%%%%%%%%%%%%%%%%%%%%%%%%%%%%%%%%%%%%%%%%%%%%%%%%%%%%%%

%%%%%%%%%%%%%%%%%%%%%%%%%%%%%%%%%%%%%%%%%%%%%%%%%%%%%%%%
% Figure 4
%%%%%%%%%%%%%%%%%%%%%%%%%%%%%%%%%%%%%%%%%%%%%%%%%%%%%%%%

%%%%%%%%%%%%%%%%%%%%%%%%%%%%%%%%%%%%%%%%%%%%%%%%%%%%%%%%
% Figure 5
%%%%%%%%%%%%%%%%%%%%%%%%%%%%%%%%%%%%%%%%%%%%%%%%%%%%%%%%

%%%%%%%%%%%%%%%%%%%%%%%%%%%%%%%%%%%%%%%%%%%%%%%%%%%%%%%%
% Figure 6
%%%%%%%%%%%%%%%%%%%%%%%%%%%%%%%%%%%%%%%%%%%%%%%%%%%%%%%%

%%%%%%%%%%%%%%%%%%%%%%%%%%%%%%%%%%%%%%%%%%%%%%%%%%%%%%%%
% Figure 8
%%%%%%%%%%%%%%%%%%%%%%%%%%%%%%%%%%%%%%%%%%%%%%%%%%%%%%%%

%%%%%%%%%%%%%%%%%%%%%%%%%%%%%%%%%%%%%%%%%%%%%%%%%%%%%%%%%%%%%%%%%%%%%%%%%%%%%%%%
% Bibliography
%%%%%%%%%%%%%%%%%%%%%%%%%%%%%%%%%%%%%%%%%%%%%%%%%%%%%%%%%%%%%%%%%%%%%%%%%%%%%%%%
\newpage
\bibliography{biblio_JPCM_CAR}
\bibliographystyle{iopart-num}
%\nocite{*}

\end{document}